\documentclass[a4paper,12pt]{article}
\pdfoutput=1
\usepackage{epsfig}
\usepackage{amssymb}
\usepackage{amsfonts}
\usepackage{amsmath}
\usepackage{euscript}
\usepackage{verbatim}
\usepackage{latexsym}
\usepackage{graphicx}
\usepackage{caption}
\usepackage{float}
\usepackage{subcaption}
\usepackage{xcolor}
\usepackage[normalem]{ulem}
\usepackage{cancel}

\usepackage{hyperref}
\hypersetup{
    colorlinks=true,
    linkcolor=black,
    citecolor=black,
    urlcolor=blue
}

\usepackage{listings}

\usepackage{booktabs}
\usepackage{appendix}
\newif\ifdtup

\catcode`\@=11

\@addtoreset{equation}{section}

\def\@normalsize{\@setsize\normalsize{15pt}\xiipt\@xiipt
\abovedisplayskip 14pt plus3pt minus3pt%
\belowdisplayskip \abovedisplayskip
\abovedisplayshortskip \z@ plus3pt%
\belowdisplayshortskip 7pt plus3.5pt minus0pt}

\def\small{\@setsize\small{13.6pt}\xipt\@xipt
\abovedisplayskip 13pt plus3pt minus3pt%
\belowdisplayskip \abovedisplayskip
\abovedisplayshortskip \z@ plus3pt%
\belowdisplayshortskip 7pt plus3.5pt minus0pt
\def\@listi{\parsep 4.5pt plus 2pt minus 1pt
     \itemsep \parsep
     \topsep 9pt plus 3pt minus 3pt}}

\relax

\catcode`@=12

\catcode`\@=11

\def\section{\@startsection{section}{1}{\z@}{3.5ex plus 1ex minus
   .2ex}{2.3ex plus .2ex}{\large\bf}}

\def\SymBoxes#1#2#3#4{\newdimen\un@t \un@t#3%
\raisebox{#1}{\rule{#2\un@t}{#4}\hskip-#2\un@t% lower horizontal
\@tempdimb\un@t \advance\@tempdimb by-#4\@tempcntb#2\relax%
\@whilenum{\@tempcntb>0}\do{%                         % #2 vertical lines
\rule{#4}{\un@t}\hskip\@tempdimb \advance\@tempcntb by\m@ne}%
\hskip-#2\un@t \rule[\un@t]{#2\un@t}{#4}%
\rule[\un@t]{#4}{#4}\hskip-#4%             % upper horizontal line
\rule{#4}{\un@t}}\hskip-#4}                % rightest vertical line
\begin{document}
%\begin{letter}{~}

%%%%%%Define some new commands and  macros
\newcommand{\beq}{\begin{equation}}
\newcommand{\eeq}{\end{equation}}
\newcommand{\bea}{\begin{eqnarray}}
\newcommand{\eea}{\end{eqnarray}}
\newcommand{\beas}{\begin{eqnarray*}}
\newcommand{\eeas}{\end{eqnarray*}}
\newcommand{\defi}{\stackrel{\rm def}{=}}
\newcommand{\non}{\nonumber}
\newcommand{\bquo}{\begin{quote}}
\newcommand{\enqu}{\end{quote}}
%%%%%%%%%%%%%%%%
\renewcommand{\(}{\beq}
\renewcommand{\)}{\eeq}
%%%%%%%%%%%%%%%%%%%%%%%%%%%%%%%%%% definitions
\def \eqn#1#2{\beq#2\label{#1}\eeq}

\def\e{\epsilon}
\def\IZ{{\mathbb Z}}
\def\IR{{\mathbb R}}
\def\IC{{\mathbb C}}
\def\IQ{{\mathbb Q}}
\def\de{\partial}
\def\Tr{ \hbox{\rm Tr}}
\def\H{ \hbox{\rm H}}
\def\HE{ \hbox{$\rm H^{even}$}}
\def\HO{ \hbox{$\rm H^{odd}$}}
\def\K{ \hbox{\rm K}}
\def\Im{ \hbox{\rm Im}}
\def\Ker{ \hbox{\rm Ker}}
\def\const{\hbox {\rm const.}}
\def\o{\over}
\def\im{\hbox{\rm Im}}
\def\re{\hbox{\rm Re}}
\def\bra{\langle}\def\ket{\rangle}
\def\Arg{\hbox {\rm Arg}}
\def\Re{\hbox {\rm Re}}
\def\Im{\hbox {\rm Im}}
\def\exo{\hbox {\rm exp}}
\def\diag{\hbox{\rm diag}}
\def\longvert{{\rule[-2mm]{0.1mm}{7mm}}\,}
\def\a{\alpha}
\def\dag{{}^{\dagger}}
\def\tq{{\widetilde q}}
\def\p{{}^{\prime}}
\def\W{W}
\def\N{{\cal N}}
\def\hsp{,\hspace{.7cm}}

\def\br{\nonumber}
\def\IZ{{\mathbb Z}}
\def\IR{{\mathbb R}}
\def\IC{{\mathbb C}}
\def\IQ{{\mathbb Q}}
\def\IP{{\mathbb P}}
\def \eqn#1#2{\beq#2\label{#1}\eeq}

\newcommand{\C}{\ensuremath{\mathbb C}}
\newcommand{\Z}{\ensuremath{\mathbb Z}}
\newcommand{\R}{\ensuremath{\mathbb R}}
\newcommand{\rp}{\ensuremath{\mathbb {RP}}}
\newcommand{\cp}{\ensuremath{\mathbb {CP}}}
\newcommand{\vac}{\ensuremath{|0\rangle}}
\newcommand{\vact}{\ensuremath{|00\rangle}                    }
\newcommand{\oc}{\ensuremath{\overline{c}}}
\newcommand{\psizero}{\psi_{0}}
\newcommand{\phizero}{\phi_{0}}
\newcommand{\hzero}{h_{0}}
\newcommand{\psiin}{\psi_{\rh}}
\newcommand{\phiin}{\phi_{\rh}}
\newcommand{\hin}{h_{\rh}}
\newcommand{\rh}{r_{h}}
\newcommand{\rb}{r_{b}}
\newcommand{\psibnd}{\psi_{0}^{b}}
\newcommand{\psibndp}{\psi_{1}^{b}}
\newcommand{\phibnd}{\phi_{0}^{b}}
\newcommand{\phibndp}{\phi_{1}^{b}}
\newcommand{\gbnd}{g_{0}^{b}}
\newcommand{\hbnd}{h_{0}^{b}}
\newcommand{\zh}{z_{h}}
\newcommand{\zb}{z_{b}}
\newcommand{\man}{\mathcal{M}}
\newcommand{\hbr}{\bar{h}}
\newcommand{\tbr}{\bar{t}}

\begin{titlepage}
%\begin{flushright} CHEP XXXXX
%ULB-TH/09-10\\
%hep-th/yymmnnn\\ \end{flushright}
%\bigskip

\def\thefootnote{\fnsymbol{footnote}}

\begin{center}
{\large
{\bf Quantum Gravity from Gluing Consistency:}} \\
\vspace{0.2cm}
{\large {\bf AdS/CFT to Spin-Foams}
}
\end{center}

%\bigskip
\begin{center}
\hspace{0.2cm} 
Chethan Krishnan$^a$\footnote{\texttt{chethan.krishnan.physics@gmail.com}}
\end{center}

\renewcommand{\thefootnote}{\arabic{footnote}}

\begin{center}
%\vspace{0.2cm}

$^a$ {Center for High Energy Physics,\\
Indian Institute of Science, Bangalore 560012, India}\\

\end{center}
\vspace{-0.15in}
\noindent

\begin{center} {\bf Abstract} \end{center}
We argue that BF theory with $SO(d+1,1)$ gauge group provides a SymTFT ``skeleton" for building AdS$_{d+1}$/CFT$_d$ pairs via gluing consistency as the bootstrap principle, loosely analogous to how compact Chern-Simons theory is a TQFT skeleton for rational CFTs. This is a non-perturbative {\em bulk} framework for quantum gravity. The bootstrap is formulated as presentation independence of physical amplitudes within a universal, defect-closed BF sewing calculus, which includes Wilson lines, codimension-two monodromy defects, and their junctions. The flat BF connection encodes the conformal geometry of the boundary, while Wilson-line path integrals construct candidate quantum-gravity microstates and charge sectors. When the bootstrapped spectrum has a stress tensor, Ward identities naturally connect it to the $B$-field, a relation we call generalized simplicity. In large-$N$, large-gap regimes admitting a gravitational EFT, this motivates the introduction of an effective notion of the $B$-field. In the Einstein limit, this becomes the ``simplicity" relation used in spin-foam models. Conventional spin-foam models try to obtain gravity by constraining BF theory. We suggest instead that BF theory is the Platonic sewing skeleton, from which theories are to be bootstrapped. Despite the topological nature of the skeleton, perturbative local bulk fields can emerge via Wilson-towers of multi-trace primaries, reproducing their one-loop determinants. We note that the BF Fourier kernel is a suggestive higher-dimensional analogue of the modular-$S$ structure. Because the bulk gluing principle in itself does not require a boundary description, the framework is conceptually broader than AdS/CFT. We sketch how de Sitter space may fit into this picture, with a natural role for observers. Our approach elevates gluing consistency to a universal bulk principle underlying theories of quantum gravity.

\vspace{1.6 cm}
\vfill

\end{titlepage}

\setcounter{footnote}{0}

\tableofcontents

\section{A Reversal of Philosophy}
\label{intro}

Approaches to quantum gravity remind one of the parable of the blind men and the elephant: different practitioners find distinct features desirable, in their favorite approaches. Perturbative string theory gives us a remarkably controlled {\em bulk} framework, but is organized around special backgrounds and only as a {\em perturbative} expansion in the string coupling \cite{Polchinski1, Polchinski2}. Via its connection to (super)gravity, string theory connects relatively easily with bulk EFT. Canonical quantization is also a {\em bulk} approach, but it is not intrinsically perturbative. Loop quantum gravity and spin-foam models were developed precisely as background-independent, {\em non-perturbative} programs \cite{AshtekarLewandowski,Baez1,Baez2,PerezSpinFoam,OritiReview}: the difficulty is that in doing this, it becomes challenging to see how dynamics (i.e., bulk EFT) can emerge. AdS/CFT gives us a beautiful\footnote{Clearly, this is \emph{my} patch of the elephant.} setting where many of these problems have satisfying answers. A holographic CFT does provide a {\em non-perturbative} definition of AdS quantum gravity \cite{MaldacenaAdSCFT, WittenAdS}. There is also strong evidence that in the semi-classical limit, it contains bulk EFT \cite{WittenAdS}. But its microscopic formulation lives on the {\em boundary}  \cite{WittenAdS}, and so there is no clear understanding of microscopic bulk ``mechanics". This is not merely a detail, the fundamental problems we wish to tackle via quantum gravity are inherently formulated in the bulk. These include the black hole information paradox, the Big Bang, and the emergent nature of metric-based dynamical spacetime itself.

Wanting the bulk and wanting a non-perturbative/background-independent definition are both noble goals, and one would like {\em not} to sacrifice either of them -- while also clarifying the connection to bulk EFT. Motivated by some of our recent work on the nature of black hole microstates in AdS$_3$/CFT$_2$ \cite{Pradipta}\footnote{See also \cite{Vishal, Rajdeep} and earlier motivating work in \cite{Vyshnav, Vaibhav1, Vaibhav2, Datar}.}, in this paper we consider the possibility that the primitive bulk structure is a \emph{gluing problem}\footnote{Sewing is often the more commonly used word when we have the boundary theory in mind, and gluing seems to be the more bulk-oriented terminology. We will use the words ``gluing" and ``sewing" interchangeably in this paper, and let the context distinguish bulk from boundary.}:  quantum data are assigned to cuts, and amplitudes obtained by cutting and gluing must agree under every admissible change of presentation. The hope is that metric gravity should emerge from this consistency/bootstrap problem, rather than by being postulated microscopically. The idea is to take a hint from Chern-Simons TQFTs, where a bulk sewing ``skeleton" can be used as a calculus for constructing distinct rational CFTs \cite{WittenJones, CIZADE}. What happens if the CFT under question is instead higher dimensional and holographic? Is there a meaningful bulk TQFT/sewing problem that one can bootstrap the CFT out of? CFT bootstrap, see e.g. \cite{Aninda} for a review, is often studied from the boundary, and usually under special circumstances. The emphasis is usually on sphere 4-point crossing or modular invariance on the torus. What if we consider bootstrap as a global sewing problem instead?

We propose that the kinematic skeleton for a CFT$_d$ is conformal BF theory in the bulk, with gauge group $SO(d+1,1)$.  BF theory has recently appeared as the spacetime SymTFT for conformal symmetry \cite{ApruzziSpacetimeSymTFT} (see also \cite{SymTFT2, MinasianSymTFT, YiNan}), without reference to bootstrap or sewing. Here we use it as the universal representation-theoretic and topological calculus relating different factorizations. The physical theory provides the spectra, pairings and matrix elements that propagate in (known) channels, and BF theory supplies the changes of presentation: equality of the resulting amplitudes is the bootstrap principle. We will not assume that conformal BF theory defines a complete TQFT on arbitrary manifolds and bordisms. What we will assume is the existence of the appropriate blocks and channel-changing kernels: the sewing/gluing calculus. We will argue that there is considerably stronger evidence for the existence of such an (in-principle fully constructible) entity. 

This approach has similarities to spin-foam models, but with crucial distinctions. Conventional gravitational spin foams also start from BF theory, but try to impose constraints on it in clever ways, to obtain gravity. Here, instead, conformal BF theory is kept fixed as the ambient kinematic sewing calculus: its blocks and change-of-presentation kernels are the universal ingredients, while the physical CFT data are to be bootstrapped from gluing consistency. This is analogous to taking the Ponsot-Teschner sewing calculus as fundamental in $d=2$, but without trying to obtain holographic CFTs by constraining Virasoro/Teichm\"uller TQFT. We expect nonetheless that the sewing calculus can bootstrap the space of CFTs.

For spherical cuts this contains the ordinary 4-point conformal bootstrap. In $d>2$, however, the same CFT amplitude can admit additional bulk presentations involving non-spherical topology. The corresponding BF block resolutions need not be purely electric: alongside Wilson-network blocks, defect-decorated fillings can provide magnetic or mixed electric-magnetic blocks, with (we conjecture) BF sewing kernels relating the different presentations. Codimension-two monodromy defects and their junctions are therefore naturally part of the same sewing calculus. The crude mental model is the TFT sewing construction of rational CFTs: a bulk topological theory gives the universal calculus of blocks and changes of presentation, while the physical CFT data determine how and which of these blocks are combined \cite{FRS1,FRS5,FFRSDefects,KongRunkelMorita}. There is one simplification in 2d, however, which is that bulk defects in the dual channel are also lines. In our higher dimensional set up instead, the natural defects are codimension-2 and are genuinely distinct from Wilson lines. A related fact is that in $d=2$, all connected cuts are circles.

The same conformal BF theory also admits an AdS interpretation on its non-degenerate Cartan branch, where the soldering component of the flat connection defines a boundary conformal geometry. For $d>2$, BF flatness implies ``normality" and restricts this geometry locally to conformal flatness. This is satisfying, because unlike in AdS$_3$ gravity \cite{AchucarroTownsend,Witten2plus1Gravity}, there is no natural Drinfel'd-Sokolov reduction \cite{BrownHenneaux,CoussaertHenneauxVanDriel} available. As in AdS$_3$, Wilson-lines become natural bulk carriers of the (local) labels selected by the bootstrap. In higher dimensions, the (conformal) charges are measured by the linking $B$-flux and not the linking holonomy. We also show that despite the topological nature of the skeleton, ordinary perturbative bulk fields are still visible when the (bootstrapped) spectrum has towers of multi-trace primaries. This follows the Wilson-tower picture proposed in AdS$_3$ \cite{Vishal}: the idea is that if the primaries that source the Wilson lines in the bulk come in a multi-trace tower, the bulk one loop determinant in thermal AdS$_{d+1}$ is simply the sum over their individual Polyakov loops. This is strong evidence that local bulk fields are a way of organizing the collective dynamics of such weakly coupled Wilson line towers\footnote{We also expect the effective dynamics of heavy microstates to be more complicated than in AdS$_3$, because the stress tensor and its multi-tower are not (Virasoro) descendants of the vacuum. Relatedly, we expect a Wilson line carrying a heavy primary label to be a heavy, charged (mass and angular momenta) state of the theory: but it is a microstate, not a black hole. When a gravitational EFT exists, a charge sector is best viewed as an ensemble \cite{Rajdeep}. Non-trivial local curvature is therefore to be viewed as an effective phenomenon associated to interactions and collective dynamics.}.

The stress tensor, when it is present in the bootstrapped spectrum, provides a bridge from kinematic structure to gravity. Its Ward identities allow the conserved conformal currents to be identified with cohomology classes of the BF $B$-field \cite{ApruzziSpacetimeSymTFT}, a relation we call \emph{generalized simplicity}.  This statement does not presuppose an Einstein dual. We argue that in a large-$N$, large-gap regime admitting a local gravitational EFT and the usual AdS/CFT correspondence, the same stress-tensor sector can in some regimes, motivate the definition of an {\em effective} $B$-field. In the Einstein gravity limit, this definition matches the so-called ``simplicity" constraint demanded in spin-foam models to reach Einstein gravity. Einstein gravity is thus a distinguished corner of the bootstrap data, while finite-gap corrections deform this relation and generate the higher-derivative gravitational EFT.  In this sense a dynamical metric and diff-invariant bulk dynamics are emergent properties of suitable gluing-consistent quantum data, that includes the stress tensor tower. 

Section~\ref{sec:BF-bootstrap} outlines the gluing calculus and the bootstrap problem. Section~\ref{sec:asymptoticads} discusses the emergence of conformal geometry on the boundary in BF theory, and the construction of bulk eigenstates of quantum gravity. Section~\ref{sec:tower} is about the Wilson-tower organization of local bulk fields, and Section~\ref{sec:ward-simplicity} develops generalized simplicity and its Einstein and adjacent limits. The concluding Comments in Section \ref{sec:conc} place these ideas in a broader context. Two directions in Section \ref{sec:conc} are worthy of emphasis: (a) We outline a relational/observer bootstrap approach to de Sitter quantum gravity (which is the topic of an upcoming paper \cite{CK}). (b) We note that the canonical BF-Fourier transform suggests a natural higher dimensional {\em bulk} generalization of the $d=2$ modular-$S$. The latter was related to a bulk Chern-Simons Fourier transform in \cite{Pradipta}. This offers a path to Cardy-like ideas in higher dimensions, which will be developed in \cite{Panikkar}.

\section{Conformal BF SymTFT and the Gluing Calculus}
\label{sec:BF-bootstrap}

A local CFT \cite{SegalCFT} is more than a collection of correlators on $\mathbb R^d$. When formulated on a specified class of Euclidean conformal manifolds, it assigns amplitudes to such $d$-manifolds and states to codimension-one cuts. Cutting a path integral prepares states on the cut, while gluing contracts those states. The familiar local conformal bootstrap is the special case in which the cut is spherical, so that radial quantization identifies its states with local operators. In dimensions $d>2$, however, one may also cut along surfaces of non-spherical topology. Sewing across such a cut involves states associated with that surface, which -- unlike the state space on $S^{d-1}$ -- has no canonical identification with the local-operator spectrum through radial quantization.

Our proposal is that conformal BF theory provides the universal\footnote{Here ``universal" refers to the conformal sewing sector. A particular CFT may carry additional internal or generalized-symmetry data, which may require additional SymTFT factors. We will ignore these in this paper.} \emph{bulk} sewing calculus organizing these different presentations. The basic object is a single sewing-closed BF calculus generated jointly by electric Wilson lines, codimension-two monodromy defects, and whatever junction sectors are required by their closure. An admissible bulk presentation consists of a filling together with a compatible bulk network structure of defects and Wilson lines realizing the prescribed asymptotic data, and defines a block basis. For fixed internal decoration labels, it defines a BF block. Cutting a block open prepares states in the corresponding BF cut state spaces, whose state-space resolutions may then be used to glue the pieces and evaluate the block. Distinct admissible presentations of the same asymptotic CFT datum give different block expansions of the same physical amplitude, and presentation independence of the resulting full amplitude is the bootstrap condition.

When the CFT itself provides an independently specified physical spectral resolution for a presentation, the corresponding spectra, pairings, matrix elements, or defect/junction data are physical CFT inputs. Other admissible BF resolutions (ie., expansions in other block bases) of the same CFT data may or may not correspond to (known) physical CFT channels. If no physical channel interpretation is known, the coefficients simply specify the representation of the amplitude in that BF presentation, without being assigned an independent CFT spectrum interpretation. Bootstrap constraints arise when the same independently defined CFT data recur in different admissible presentations and must be represented consistently in all of them. For the familiar spherical four-point decompositions this reduces to ordinary conformal crossing entirely within the Wilson-line subsector. On more general topologies sewing closure may require bulk monodromy defects and the full junction sectors. The problem is therefore to determine which physical CFT data are compatible with presentation independence in this universal defect-closed BF calculus. The picture we present has broad analogies with the TFT construction of distinct RCFTs \cite{FRS1,FRS5,FFRSDefects}.

We will ignore possible anomaly contributions, and also restrict attention to conformally flat asymptotic boundary geometries. Our goal is simply to present what we believe is the simplest non-trivial version of the construction. When present, anomalies should be incorporated through additional Chern-Simons or invertible sectors of the spacetime SymTFT \cite{ApruzziSpacetimeSymTFT}. Ignoring them means that some of our statements may apply only in odd $d$, but we expect the basic philosophy to apply in even $d$ as well. Conformal flatness of the boundary on the other hand, seems to be preferred by the bulk BF theory, as we will see in the next section.

\subsection{The Gluing Problem in Higher Dimensions}
\label{subsec:BF-sewing-problem}

Let $X^d$ be a Euclidean CFT spacetime and let $C^{d-1}\subset X$ be an oriented codimension-one cut. Writing
\beq
 X=X_L\cup_C X_R ,
\label{eq:CFT-cut-decomposition}
\eeq
the path integrals on $X_L$ and $X_R$ prepare states
\beq
 |\Psi_L\rangle\in {\cal H}_{\rm CFT}(C),
 \qquad
 \langle\Psi_R|\in {\cal H}_{\rm CFT}(\bar C)
 \simeq {\cal H}_{\rm CFT}(C)^* ,
\label{eq:CFT-cut-states}
\eeq
and gluing along $C$ gives the pairing
\beq
 Z[X]=\langle\Psi_R|\Psi_L\rangle_C .
\label{eq:CFT-gluing}
\eeq
When the CFT has an independently specified channel decomposition for a cut -- a case we will call a \emph{physical channel} -- sewing may be expressed by resolving the identity on ${\cal H}_{\rm CFT}(C)$ into those physical states and summing or integrating over them. Local operators in radial quantization provide such a basis for sphere cuts. We will not assume {\em a-priori} that an arbitrary cut comes with a known physical spectral decomposition. But sewing must happen nonetheless: by the path-integral gluing law, any CFT consistently defined on the admitted backgrounds provides the corresponding state spaces and amplitudes, and hence admits resolutions appropriate for gluing. Our proposal is that these sewing resolutions can be realized kinematically by the corresponding BF state spaces and kernels.

For the spherical cut
\beq
 C=S^{d-1},
\eeq
radial quantization gives the state-operator correspondence. A complete set of channel states is therefore organized by local conformal multiplets, and the familiar local data are
\beq
 (\Delta_i,\rho_i),\qquad
 C_{ijk}^{\alpha},\qquad
 G_{ij},\qquad
 {\bf 1},\qquad
 T_{\mu\nu},\qquad C_T,
\label{eq:BF-local-physical-data}
\eeq
where $\rho_i$ denotes the Lorentz representation, $\alpha$ labels independent tensor structures and $G_{ij}$ is the non-degenerate two-point pairing. Four-point crossing on $S^d$ is the familiar presentation-independence condition in this local sector.

The situation changes in higher dimensions because $C$ need not be a sphere. For example, we expect that a three-dimensional CFT can be quantized on
\beq
 S^2,\qquad T^2,\qquad \Sigma_g,\ldots
\eeq
Each of these can be realized as a spatial slice of an appropriate conformally flat background\footnote{Very crudely, the philosophy is that {\em local} quantum field theory is defined locally, and therefore we expect to be able to patch it up to live on arbitrary manifolds. This is essentially another articulation of the idea of sewing. There can be global obstructions, and it is not clear to me whether a full set of manifolds on which a general CFT can be defined, has been conclusively charted. There is also a second question, about what kind of background conformal classes we should allow. Because of the presence of the stress tensor, one expects to be able to couple the CFT to arbitrary backgrounds. We will assume that the backgrounds are conformally flat. This includes all manifolds in $d=2$ and quite a general class (M\"obius manifolds) in $d>2$ including the boundary AdS$_{d+1}$ cylinder (and the ones we have listed above). This class may need refinement, as we better understand the bordisms and manifolds on which BF sewing calculus is well-defined. We will take the bulk to be the eventual arbiter in this matter, in deciding the class of boundary manifolds under consideration.}. The Hilbert spaces ${\cal H}_{\rm CFT}(T^2)$ or ${\cal H}_{\rm CFT}(\Sigma_g)$ are QFT state spaces, but there is no analogue of the local state-operator map which identifies every such state with an operator inserted at a point. The torus state-operator correspondence has in fact been studied directly and obstructions noted \cite{BelinTorus}. The existence of non-spherical cuts therefore makes the problem of identifying the preferred physical spectral decomposition, challenging.

This does \emph{not} mean that each topology defines a new theory. The same full CFT assigns all of these Hilbert spaces, so the corresponding data must fit together\footnote{While we expect {\em consistency}, we are not claiming that the higher-topology data are determined {\em uniquely} by the local spectrum and OPE data. While the local bootstrap conditions are very stringent, we are not aware of a uniqueness statement about the full theory, from only local data.}. \emph{But} in $d>2$ there is no known analogue of the two-dimensional sewing results showing reconstruction of all higher-topology amplitudes \cite{SonodaSewingI,SonodaSewingII}.  

The claim that we will elaborate below, is that in higher dimensions, once an amplitude is fixed by physical data in one channel, the BF sewing calculus represents that same amplitude in other admitted BF block resolutions. The resulting coefficient distribution is simply its representation in the new BF block basis: it may not be supported as an independently known/defined physical spectrum of the CFT in that channel. 

We expect that BF quantization (e.g., with defects) is a natural way to characterize states in such channels\footnote{E.g.: we expect that defect-inserted bulk BF path integrals define generalizations of Wilson line characters (which are associated to local spectra, see Section \ref{sec:tower}), to non-spherical topology.}. Constraints can arise when the same BF-represented data recur in different presentations and must agree. In particular, when another presentation has its own physical channel interpretation, the transported coefficients must agree with its independently defined CFT data. Thermal and higher-dimensional modular consistency conditions give potential examples in which distinct physical presentations of the same CFT may provide useful constraints \cite{ShaghoulianGeneralizedCardy,HorowitzShaghoulianDetachable,IliesiuThermalBootstrap}. But of course, the most famous and concrete example is the 4-point sphere case in two different presentations \cite{Rychkov}.

There is a second (and complementary) enlargement which is part of the BF formulation. Once the bulk sewing calculus contains topological defects, a complete BF resolution need not remain inside the Wilson-line-generated subspace. A general BF block basis\footnote{We remind the reader that a block basis, is what we also call a presentation.} for a given CFT amplitude need not be defined only by a Wilson network fusion spine, but may also contain defect decorations. 

We package the independently physical CFT input as data\footnote{When there is no room for confusion, sometimes we will also refer to it loosely as the spectrum, even though it can contain OPE and other data.}, to be eventually bootstrapped:
\beq
{\mathsf D}
=
\left\{
\begin{array}{l}
\text{Physical spectra, pairings and matrix elements, including defect/junction data}\\
\text{in channels with an independent CFT interpretation, }\\
\text{reused whenever such a channel recurs.}
\end{array}
\right\}.
\label{eq:BF-sewing-data-general}
\eeq
Generic BF coefficient distributions are induced by representing the physical amplitude in the corresponding BF resolution. We expect that a complete definition of CFT physical data in a presentation is to be characterized by quantizing BF theory with the appropriate lines and defects. The simplest examples are the characters obtained via bulk Wilson lines in BF theory, relevant for local operator sectors. BF theory can compute characters (the simplest analogues of blocks) given the label, but the spectrum of Wilson lines in a given CFT is data to be bootstrapped. With more general defects, we expect monodromy data, centralizers, etc. to play a role in determining the BF quantization.

The restriction of ${\mathsf D}$ to spherical Wilson-line channels is precisely the ordinary local spectrum and OPE data in \eqref{eq:BF-local-physical-data}. The local conformal bootstrap is therefore a distinguished subsystem of the full sewing problem.

As an aside, let us mention the simplification in $d=2$. All manifolds are conformally flat, and organized as Riemann surfaces. A connected closed CFT cut is a circle, so arbitrary surface sewings reuse the same circle state space. The topology of the worldsheet can be built by sewing circles, there is no other connected cut topology. This is one reason why the Moore-Seiberg/RCFT sewing calculus can be organized in terms of local sectors. In $d>2$ there are genuinely different connected cut topologies\footnote{There is also the technical difference that the sewing calculus in $d=2$ for a generic CFT is Virasoro/Ponsot-Teschner and not BF, but that is only a technicality for the point we are making here.}, and the possibility that their natural BF resolutions involve different kinds of topological data is part of the bootstrap problem. 

\subsection{Conformal BF Theory and the Defect-Closed Gluing Calculus}
\label{subsec:BF-sewing-calculus}
\label{subsec:BF-electric-magnetic-completion}

For a Euclidean $d$-dimensional CFT we take
\beq
 G=SO(d+1,1).
\label{eq:BF-conformal-group-bootstrap}
\eeq
The choice of connected component or cover can be suitably accommodated, and we will not emphasize it. The local $(d+1)$-dimensional BF theory\footnote{We review BF theory and its connection to canonical gravity more explicitly in Appendix \ref{app:BF}. But our discussion here is more abstract, and should not require the details.} depends only on the Lie algebra, while the labels of extended operators can depend on the global form of the gauge group:
\beq
 S_{\rm BF}[A,B]
 =
 \frac{1}{\kappa_{\rm BF}}
 \int_{M^{d+1}}\!\langle B\wedge F[A]\rangle .
\label{eq:BF-bootstrap-action}
\eeq
Here $A$ is a $G$ connection and $B$ an adjoint-valued $(d-1)$-form \cite{HorowitzBF,BlauThompsonBF,Baez4dBF}. Away from sources,
\beq
 F[A]=0,
 \qquad
 D_A B=0.
\label{eq:BF-bootstrap-eom}
\eeq
The conformal spacetime SymTFT of \cite{ApruzziSpacetimeSymTFT} contains precisely such a BF sector, but without attempting to do sewing or bootstrap. The manifold $M^{d+1}$ used here is not being identified with the auxiliary interval direction of the SymTFT sandwich construction \cite{ApruzziSpacetimeSymTFT}. We use it instead as a \emph{sewing skeleton}: BF theory is taken to provide the universal (i.e., CFT-independent) maps which compare presentations, but it does not by itself include -- except in the Platonic sense that all bootstrapped CFTs are solutions of it -- the physical spectra, OPE coefficients, or defect data. We will view BF theory as a bulk ``TQFT"\footnote{We will not demand that this TQFT is well-defined on the full bordism category. But we expect that it is well-defined enough on a subset (loosely in analogy with Virasoro/Teichm\"uller TQFT), that one can use its sewing calculus to bootstrap CFTs, as we will make more precise below. We will use the phrases ``sewing/gluing skeleton", ``TQFT skeleton", ``sewing/gluing calculus" or even simply ``BF TQFT", with this meaning.}. The gauge symmetry of the bulk TQFT becomes the global conformal symmetry of the boundary, in the spirit of the SymTFT language.

One basic class of sources is formed by electric Wilson lines
\beq
 W_R(\gamma)
 =
 {\cal P}\exp\!\left(\int_\gamma A\right)_R ,
\label{eq:BF-electric-line-bootstrap}
\eeq
labelled by representations $R$ of $G$. An open Wilson line carries a conformal representation label, so it is natural to associate the boundary endpoints of open Wilson lines to local operator insertions. This remains true even when the ambient boundary is not a sphere, because one may remove a small sphere around the insertion and perform radial quantization on that sphere. Thus the CFT labels $(\Delta,J,\ldots)$ are the boundary representation labels carried by these electric lines. We will also use analytic continuation in these labels, as often done in the CFT bootstrap literature. 

A second basic class consists of codimension-two monodromy defects \cite{BaezPerezBFBranes}. Locally, a defect worldvolume $\Sigma_m^{d-1}\subset M^{d+1}$ is characterized by a prescribed conjugacy class of meridional holonomy,
\beq
 \operatorname{Hol}_{\mu}(A)\in[g_m],
\label{eq:BF-magnetic-holonomy-bootstrap}
\eeq
where $\mu$ is a small loop linking $\Sigma_m$. More generally, a monodromy defect may also carry electric data associated with the centralizer of its magnetic holonomy\footnote{Such mixed electric-magnetic defects are sometimes called dyonic, but to avoid clutter, we will often use ``defects" to include everything.}. Wilson lines, defects, and their allowed junctions and fusion channels are ingredients of one BF sewing calculus. We will denote by $\mathfrak d$ the actual topological decoration data carried by a bordism, including Wilson lines with their representation labels, monodromy defects with their prescribed conjugacy classes and any additional electric data, and junction data when present.

In usual discussions of sewing and bootstrap, cutting and gluing is the language through which block basis changes are implemented. We find it more convenient to treat a block basis itself as more fundamental (this is what we call a presentation), and view the cut-and-glue operation as being implemented via BF kernels that implement such presentation changes\footnote{These are sometimes also called channel changes (in the sense of $s$ and $t$ channels), and even though we do occasionally use the word ``channel" when there is no possibility of confusion, we will generally stick with presentation change or block basis change, when we want to be precise. We will use channel as a broad name for a (sometimes internal) basis, and  also the physical CFT data as realized by the amplitude in a {\em given} presentation.}. 

For an admissible decorated filling $M^{d+1}$, if $C^{d-1}\subset X^d=\partial_{\rm as}M$ is a boundary sewing surface, one may introduce a $d$-dimensional BF cut $Y^d\subset M$ whose asymptotic boundary contains $C$,
\beq
 \partial Y\cap\partial_{\rm as}M=C .
\label{eq:BF-boundary-bulk-cut}
\eeq
If $Y$ meets an electric Wilson graph $\Gamma_E$ transversely and away from its junctions, the cut inherits marked points $P=Y\cap\Gamma_E$ with the corresponding representation labels. If it meets a magnetic worldvolume, it inherits a codimension-two locus $L=Y\cap\Sigma_m$ around which the cut connection has the specified meridional holonomy. More generally, the cut carries the induced decoration $\mathfrak d_Y$ obtained by restricting the bulk decoration $\mathfrak d$ to $Y$. BF quantization is then assumed to assign to the decorated cut, with its induced asymptotic boundary conditions understood, a kinematic state space
\beq
 {\cal H}_{\rm BF}(Y;\mathfrak d_Y) .
\label{eq:BF-defect-cut-state-space}
\eeq
Roughly: the bordisms on either side of $Y$ prepare states in this state space. Their topology and decoration go into the determination of those states, rather than the definition of the cut state space itself. For a Wilson-only cut, the induced decoration $\mathfrak d_Y$ consists of  marked points and representation labels. More generally, we can think of $\mathfrak d_Y$ as the part of the bulk decoration that intersects the cut, and the corresponding state space is obtained by BF quantization with this induced decoration\footnote{There are precedents for defect-inserted BF quantization, including the membrane spin-network construction of \cite{BaezPerezBFBranes,FairbairnPerezBF}. But a sufficiently extensive defect-BF quantization for the conformal group does not appear to have been developed.}. Even an undecorated cut with non-contractible cycles may carry global BF degrees of freedom. In the connection polarization used here, these are represented by flat-connection holonomies around the non-contractible cycles when they are not already fixed by the asymptotic boundary conditions: the conjugate $B$-field degrees of freedom are implicit in the BF quantization. Holonomies fixed by the boundary conditions are fixed parameters, rather than degrees of freedom on $Y$.

Geometrically, we assume a sewing-closed class $\mathfrak C$ of fillings, cuts, and bordisms compatible with the asymptotic boundary data. We will call a filling, cut, or bordism \emph{admissible} when its underlying geometry belongs to $\mathfrak C$. Admissible bordisms support the decorations of the defect-closed BF calculus, with the corresponding induced data on their cuts, and BF quantization assigns the associated state spaces, amplitudes, pairings, gluing operations, and change-of-presentation maps. Individual bordisms may be represented by Morse decompositions into elementary handle attachments $D^k\times D^{d+1-k}$, but we do not identify $\mathfrak C$ with the full category of $(d+1)$-dimensional BF bordisms, nor require such a full TQFT construction here.

The (simpler) mental model for this is the $d=2$ Ponsot-Teschner/Moore-Seiberg sewing calculus \cite{MooreSeibergCFT,PonsotTeschner1,TeschnerTeichmuller}. Intrinsically, that calculus is formulated in terms of punctured Riemann surfaces and their different pants decompositions or markings\footnote{A pants decomposition is the three-holed spheres (pairs of pants) into which a  Riemann surface is cut up. A marking includes, in addition, the associated trivalent Moore-Seiberg graph, which carries the combinatorial ``spine'' information needed to specify how the pairs of pants are assembled. In other words, markings specify a conformal-block basis. In a bulk TFT realization this graph is naturally represented by a Wilson or ribbon network. It is this block-basis data that we will call a presentation (in higher dimensions).}: the relevant structure is the class of surfaces, cuts, and elementary changes of presentation needed for sewing, rather than the category of all three-dimensional bordisms. In a bulk language, these sewing operations can be realized by a corresponding class of three-dimensional fillings and bordisms. Different markings of the same boundary surface, and correspondingly different filling/network presentations in a bulk realization, provide different conformal-block bases. In our higher-dimensional generalization, we will allow the filling to be part of the presentation whenever it is needed to specify the BF-block basis and the sewing calculus relates it to other admissible presentations.

We will regard this higher-dimensional sewing domain $\mathfrak C$\footnote{The basic extra complication in higher dimensions is that spheres do not exhaust the possible cut topologies. A cut with non-contractible cycles can carry global flat-connection degrees of freedom when these are not fixed by the asymptotic boundary data, together with any decoration induced from the bulk defect network. Both are incorporated in the BF quantization of the corresponding cut state space. Later in this section we review circumstantial evidence from BF theories with other gauge groups that such global quantization and sewing are possible.} as the analogue of the domain underlying the Moore-Seiberg sewing problem. Unlike Moore-Seiberg, we do not (necessarily) assume the existence of a finite generating set of higher-dimensional sewing moves, even though we suspect it does. A more intrinsic or maximal geometric characterization of this domain is a mathematical problem for the future. We will view it roughly as analogous to passing from the Ponsot-Teschner sewing to the Teichm\"uller/Virasoro-TQFT.

This then, is the central structural hypothesis of the construction: on an appropriate sewing-closed domain $\mathfrak C$, conformal BF theory admits a coherent defect-closed sewing calculus, with state spaces, gluing operations and change-of-presentation kernels, and this calculus is sufficiently rich to provide generalized BF resolutions of the CFT amplitudes under consideration and to connect the recurring physical channels whose compatibility is to be bootstrapped. We do not assume that such a calculus exists on arbitrary manifolds and bordisms. The spherical sector below realizes this sewing structure explicitly. The thermal-AdS Wilson-character resolution of Section~\ref{subsec:thermal-wilson-character} is suggestive that the BF presentation calculus  extends beyond the local spherical four-point setup. For higher-topologies, in a later subsection, we will present various plausibility arguments, assembled and built from the literature.

The datum held fixed by the bootstrap is the asymptotic CFT configuration that captures a correlator, partition function, or more generally, an amplitude:
\beq
 \mathfrak X=(X^d,\mathfrak d_\partial),
\label{eq:BF-asymptotic-datum}
\eeq
where $\mathfrak d_\partial$ denotes the prescribed asymptotic decoration, including the local insertions or external (CFT) defect data when present. A \emph{presentation} $\mathfrak p$ of $\mathfrak X$ is the bulk realization of a BF block basis: it consists of an admissible filling $M^{d+1}$ with $\partial_{\rm as}M=X$, together with a compatible family of bulk decorations $\mathfrak d_\lambda$ realizing $\mathfrak d_\partial$, and the network structure specifying the corresponding block basis. The individual blocks in this basis are labelled by the internal BF channel data $\lambda$. 

We will view the block basis (presentation) as the fundamental object. The basis independence of the amplitude, is what ultimately leads to bootstrap.

For fixed $\mathfrak p$ and fixed internal channel labels $\lambda$, the corresponding decorated filling $(M,\mathfrak d_\lambda)$ computes the kinematic BF block $\Psi^{\rm BF}_{\mathfrak p,\lambda}$. The presentation $\mathfrak p$ thus determines a generalized BF block basis for the boundary datum $\mathfrak X$, which we write as\footnote{The standard RCFT four-point problem provides a map for this terminology. Consider four local operators on the sphere in the $s$-channel fusion/network presentation. A circle separating $(1,2)$ from $(3,4)$ exposes the corresponding intermediate channel, which can contain more than one irreducible intermediate sector. In a multiplicity-free RCFT, the blocks in this resolution are labeled simply by the allowed intermediate representations $p$. In a general RCFT one must also choose bases in any fusion-multiplicity spaces. Holding $p$ fixed gives the corresponding $s$-channel conformal block $\mathcal F^{(s)}_p$ (with the descendant states inside the representation already incorporated into the block). $\Psi^{\rm BF}_{\mathfrak p,\lambda}$ is the BF analogue of the block. For this four-point example the bulk filling is held fixed, so the non-trivial presentation data are the fusion/network realization and the resulting block basis, while the circles (cuts) provides a convenient way to expose its intermediate channel. In general, the filling is also part of $\mathfrak p$. The intermediate label $p$ is the analogue of $\lambda$. A different presentation, for example the $t$-channel fusion/network, gives a different basis $\mathcal \{F^{(t)}_q\}$, and the two bases are related by the RCFT $F$ matrix. The analogy is only at the level of the kinematic block resolution. In a full RCFT the coefficients are fixed by the physical CFT data (which can be viewed as bootstrapped). Likewise, in our construction the physical CFT amplitude is represented in each admitted BF resolution by a coefficient distribution determined by the same underlying CFT data. See \eqref{eq:BF-channel-amplitude-bootstrap}.}
\begin{equation}
\big\{\Psi^{\rm BF}_{\mathfrak p,\lambda}\big\},
\qquad
\lambda\in\Lambda_{\mathfrak p},
\qquad
d\mu_{\mathfrak p}(\lambda),
\label{eq:BF-block-basis-bootstrap}
\end{equation}
where $\lambda$ collectively denotes the internal BF data required to label the blocks in that presentation, and $d\mu_{\mathfrak p}(\lambda)$ denotes the corresponding generalized basis measure. In the spherical Wilson-line presentations considered explicitly below, these include representation, multiplicity, and tensor-structure labels, with the principal-series representation labels summed and integrated using the appropriate conformal Plancherel measure. In more general presentations the required labels are determined by the corresponding defect-closed BF sewing calculus.

A second presentation $\mathfrak p'$ of the same boundary datum $\mathfrak X$ may differ in its filling and/or in its compatible bulk network realization, and hence in the BF block basis it defines. The BF sewing calculus relates the corresponding kinematic block bases by a change-of-presentation kernel
\beq
\Psi^{\rm BF}_{\mathfrak p,\lambda}
 =
 \int d\mu_{\mathfrak p'}(\lambda')\,
 K^{\rm BF}_{\mathfrak p\to\mathfrak p'}
 (\lambda,\lambda';\mathfrak X)\,
 \Psi^{\rm BF}_{\mathfrak p',\lambda'} .
\label{eq:BF-presentation-kernel-bootstrap}
\eeq
These blocks and change-of-presentation kernels are the kinematic sewing ingredients needed by the bootstrap. We expect that a general $K^{\rm BF}_{\mathfrak p\to\mathfrak p'}$ can mix a Wilson-spine-generated basis with one that includes defects.

The analogy with Ponsot-Teschner sewing in $d=2$ should be clear. We view Teichm\"uller/Virasoro TQFT as the bulk gluing realization of that calculus. Our BF gluing skeleton is a rough higher-$d$ analog of the bulk gluing. The key difference (loosely) in $d=2$ is that because all cuts are circles, all channels have a straightforward ``physical" understanding within the CFT via radial quantization. 

A simple illustration of the parallel is provided by torus characters in the dual channel\footnote{A channel in the sense which it is used in this context, is identical to what we are calling a presentation.}. One could view them as being computed via ``defect" insertions: in $d=2$, these are again Wilson lines, but placed at the core of a contractible {\em thermal} cycle in the dual filling, and running along the (non-contractible) spatial cycle. The sewing calculus relates this ``defect" block basis element to the direct channel basis block element (again a character, but with modulus $\tau$ instead of $-1/\tau$) via the modular-$S$ kernel. Of course, in the $d=2$ case one never views these as defect block basis elements: in $d=2$ we have the luxury of modular invariance to understand the relation between the two channels. Our claim is that presentation change is the more fundamental perspective, in that it is amenable to generalization to higher dimensions. We will see later that similar basis changes may be natural in the $S^2 \times S^1$ geometry in $d=3$ as well, depending on whether we choose to view the $S^1$ as contractible in the bulk, or not.

Another apparent difference with $d=2$, is the ``obvious" fact that the bulk seems less topological in higher dimensions. This will be explained later via the presence of an independent stress tensor current (and multi-stress-tensor composites) in the bootstrapped spectrum of primaries. An interesting question in higher dimensions is whether one can define a natural analogue of Liouville CFT from these sewing ingredients. We will discuss that question in upcoming work.

\subsection{Basic Consistency: Crossing on the Sphere from BF TQFT}

The simplest non-trivial test for the existence of the sewing skeleton above is the four-point spherical channel. It is useful to formulate the relevant change of presentation first entirely in a bulk BF theory\footnote{Our bulk discussion here applies to fairly general BF gauge groups, and is not limited to conformal BF theory.}, before comparing with CFT harmonic-analysis results. Even though the material is supposed to be standard, we outline the story in some detail, because the two ends are standard in two different communities.

Take a bulk presentation corresponding to a spherical four-point channel. The local change from one channel to another can be isolated in a contractible bulk region containing a trivalent Wilson tree with four external edges labelled by representations $R_1,\ldots,R_4$ of $G$.  A Wilson insertion depends only on $A$, so the $B$ equation of motion continues to impose flatness $F[A]=0$, away from magnetic defects. We can therefore trivialize the connection on this region. In such a trivialization a Wilson segment from $y$ to $x$ is simply a representation matrix,
\bea
W_R(x,y)
\equiv
{\cal P}\exp\!\left(\int_y^x A_R\right)
=
R\!\left(g(x)g^{-1}(y)\right), \label{Wilson-Rep}
\eea
where $A_R$ denotes the connection acting in $R$. At a trivalent junction the incident representation indices are contracted with an intertwiner. The intertwining property cancels the group element associated with the internal vertex, so the dependence on its position disappears\footnote{An intertwiner is a ``$G$-equivariant map" between representation spaces. For a trivalent vertex, it is a $G$-Clebsch-Gordan tensor, $I:V_{R_1}\otimes V_{R_2}\to V_{R_3}$ with $I\,[R_1(g)\otimes R_2(g)]=R_3(g)\,I$. Dualizing the outgoing leg, one can also write this as $I \in \operatorname{Inv}_G\!\left(V_{R_1}\otimes V_{R_2}\otimes V_{R_3}^{*}\right)$. Equivalently, it is the invariant tensor that joins the representation indices of the Wilson lines meeting at the vertex.  The group element $g(z)$ at an internal vertex therefore acts simultaneously on all incident legs, and the intertwining relation moves this common action through the vertex, where it cancels against the inverse group element on the adjacent Wilson line.  Hence the network is independent of the local group element $g(z)$.}. Thus, in this elementary Wilson-line sector, a flat BF Wilson tree reduces to a composition of $G$-intertwiners: its internal structure is purely a problem in the representation theory of $G$.

This immediately gives the elementary BF change of channel.  Let $V_{R_i}$ denote the representation space carried by the $i$th external Wilson line.  A gauge-invariant four-valent Wilson vertex is an invariant tensor in the product of these four representation spaces.  Thus the space of possible four-line couplings is
\beq
 {\cal V}_{1234}
 =
 {\rm Inv}_G\!\left(
 V_{R_1}\otimes V_{R_2}\otimes V_{R_3}\otimes V_{R_4}
 \right).
\label{eq:BF-four-leg-invariant-space}
\eeq
${\rm Inv}_G$ denotes the subspace left invariant by the simultaneous action of $G$ on all four factors. Equivalently, an element of ${\cal V}_{1234}$ is a way of contracting the four representation indices to a $G$-singlet.

One convenient basis of this space is obtained by resolving the four-valent coupling into two trivalent ones\footnote{The fact that every four-line invariant can be resolved through an intermediate representation is a completeness statement for the pairwise tensor-product decomposition. The basic input is $V_{R_1}\otimes V_{R_2}\simeq \bigoplus_R N_{12}^{R}\otimes V_R$, with $N_{12}^{R}$ the multiplicity space, and hence
${\rm Inv}_G(V_{R_1}\otimes V_{R_2}\otimes V_{R_3}\otimes V_{R_4})
\simeq
\bigoplus_R
N_{12}^{R}\otimes
{\rm Inv}_G(V_R\otimes V_{R_3}\otimes V_{R_4})$.
Thus there are no additional singlets that arise ``suddenly" at four points. Once the $R_1\otimes R_2$ product has been decomposed completely into irreducibles, every four-point singlet necessarily factors through one of those intermediate sectors. For noncompact $G$, such as $SO(d+1,1)$, the direct sum is replaced by the appropriate direct integral, possibly supplemented by discrete contributions or contour prescriptions. In the conformal application below, this recoupling calculus is applied to the Euclidean induced representations of harmonic analysis, with external representation parameters analytically continued away from the principal series when needed. These should not be confused with the positive-energy lowest-weight modules ${\cal V}_{\Delta,J}$ used in radial quantization in the next subsection.
}.  In the $s$-channel, the first two external representations are coupled to an intermediate representation $R$, which is then coupled to $R_3$ and $R_4$.  The two trivalent vertices are described by invariant tensors
\beq
 C^{\alpha}_{12R^\vee}
 \in {\rm Inv}_G\!\left(
 V_{R_1}\otimes V_{R_2}\otimes V_{R^\vee}
 \right),
 \qquad
 C^{\beta}_{R34}
 \in {\rm Inv}_G\!\left(
 V_R\otimes V_{R_3}\otimes V_{R_4}
 \right).
\label{eq:BF-s-channel-intertwiners}
\eeq
Here $R^\vee$ denotes the dual representation, required because the internal Wilson line is outgoing from one trivalent vertex and incoming to the other.  The labels $\alpha$ and $\beta$ distinguish independent invariant tensors when the corresponding singlet occurs with multiplicity greater than one.  Contracting the $R^\vee$ and $R$ indices along the internal Wilson line produces an element of ${\cal V}_{1234}$. We denote the resulting $s$-channel invariant by
\beq
 \mathcal I^{(s)}_{R,\alpha\beta}
 \equiv
 C^{\alpha}_{12R^\vee}\circ C^{\beta}_{R34}
 \in {\cal V}_{1234},
\label{eq:BF-s-channel-basis}
\eeq
where the composition symbol denotes contraction of the $V_{R^\vee}$ and $V_R$ indices using their natural invariant pairing. Varying $R$ and the multiplicity labels $\alpha,\beta$ therefore gives the $s$-channel resolution of the four-line invariant tensor space. The $t$-channel similarly gives a basis $\mathcal I^{(t)}_{R',\gamma\delta}$ by first coupling a different pair of external representations.  For a non-compact group the intermediate labels are in general summed and/or integrated with the measure appropriate to the corresponding tensor-product decomposition.  The two resolutions are related by a change of basis:
\beq
 \mathcal I^{(s)}_{R,\alpha\beta}
 =
 \int d\mu_t(R')\,
 \sum_{\gamma,\delta}
 {\mathsf F}^{\;R',\gamma\delta}_{R,\alpha\beta}[R_i]\,
 \mathcal I^{(t)}_{R',\gamma\delta}.
\label{eq:BF-Racah-recoupling}
\eeq
The kernel ${\mathsf F}$ is the Racah-Wigner coefficient, or $6j$ symbol, of $G$.  In other words, for this elementary four-line move the BF change-of-presentation kernel in \eqref{eq:BF-presentation-kernel-bootstrap} is nothing but the representation-theoretic recoupling kernel. Note that while familiar from CFTs, we have reached this statement without the boundary CFT (and the statement applies to general $G$): it follows from BF flatness, Wilson lines, and the composition of intertwiners.

The corresponding coherence relation is again essentially group/representation theory. For five external representations one may pass between two binary fusion trees by different sequences of elementary recouplings.  Associativity of the tensor product requires the resulting maps to agree.  Written in terms of the kernels in \eqref{eq:BF-Racah-recoupling}, this is called the Biedenharn-Elliott, or pentagon, identity, with sums replaced by the appropriate integrals in continuous cases. Thus the elementary Wilson-line BF problem has the characteristic sequence
\beq
\begin{gathered}
 \text{flat Wilson networks}
 \quad\longrightarrow\quad
 \text{$G$-intertwiners}
 \quad\longrightarrow\quad
 \text{$G$-Racah kernels}
 \\[-2pt]
 \longrightarrow\quad
 \text{pentagon coherence}.
\end{gathered}
\label{eq:BF-wilson-evidence-chain}
\eeq
This is the local recoupling structure that the sewing proposal of the previous subsection requires in the four-point spherical sector.

We now briefly outline the spherical cut basis for the conformal BF sewing calculus \cite{DobrevMackHarmonic,KarateevHarmonic}.  Take $G=SO_0(d+1,1)$, with the choice of cover (for allowing fermionic representations) understood as mentioned earlier.  For the spherical cuts relevant to the local four-point problem, we choose the BF channel basis to be the Plancherel basis of Euclidean conformal harmonic analysis.  The intermediate labels are then the unitary principal-series representations
\beq
 R=(\Delta,\rho),
 \qquad
 \Delta=\frac d2+i\nu,
 \qquad \nu\in\mathbb R,
\label{eq:BF-principal-series-bootstrap}
\eeq
where $\rho$ denotes the $SO(d)$ spin representation.  This resolution comes equipped with invariant pairings, the Plancherel measure, and the shadow transform\footnote{The shadow transform is a conformally equivariant linear map, i.e.\ an intertwining operator between a representation and its shadow, $V_{\Delta,\rho}\to V_{d-\Delta,\rho^R}$, where $\rho^R$ is obtained from $\rho$ as a parity-reflected $SO(d)$ representation. In reflection-invariant spin sectors, including symmetric traceless tensors, $\rho^R\simeq\rho$.  On the principal series, $d-\Delta=\Delta^*$, and the representation and its shadow are related by the invariant shadow pairing. In position space this shadow pairing is what appears when two three-point intertwiners are sewn together, producing the single-valued conformal partial wave below.} \cite{DobrevMackHarmonic,KarateevHarmonic}. Accordingly, the BF completeness relations and change-of-channel integrals on these cuts are taken over the principal-series contour (i.e., the real line parametrized by $\nu$).  This is the sewing resolution that we will take for the TQFT skeleton (which should not be confused with the CFT spectrum). For physical correlators, the corresponding generalized harmonic expansion may also require isolated contributions not represented by the principal-series integral itself, with the identity operator providing the basic example.

To connect this abstract representation-theoretic description with the more familiar CFT language, consider the standard position-space realization of the conformal-group representations.  In this realization the action of $G$ on a representation $V_R$ is represented by conformal transformations of a field ${\cal O}_R(x)$.  Consequently, the statement that a tensor is a $G$-singlet becomes a conformal Ward identity.  For example, the condition
$C_{123}\in{\rm Inv}_G(V_{R_1}\otimes V_{R_2}\otimes V_{R_3})$
becomes schematically
\beq
 \left(\mathcal J_A^{(1)}+\mathcal J_A^{(2)}+\mathcal J_A^{(3)}\right)
 \langle {\cal O}_1(x_1){\cal O}_2(x_2){\cal O}_3(x_3)\rangle=0 ,
\eeq
where $\mathcal J_A^{(i)}$ is the conformal generator acting on the $i$th insertion.  Thus a three-valent $G$-intertwiner is realized in position space as a conformally invariant three-point tensor structure.  The non-trivial dependence on the insertion points is simply how the abstract singlet condition is represented in this basis.

The contraction of the internal representation indices in \eqref{eq:BF-s-channel-basis} also has a familiar position-space realization. On the principal-series contour the representation paired invariantly with $R=(\Delta,\rho)$ may be represented by its shadow $\widetilde R=(d-\Delta,\rho^R)$.  The abstract contraction of the two three-valent intertwiners then becomes an integral over their common position, together with the appropriate spin-index contractions:
\beq
 \Psi^{(s)}_{R,\alpha\beta}(x_i)
 \;\propto\;
 \int d^d x\,
 \langle {\cal O}_1(x_1){\cal O}_2(x_2){\cal O}_{R}(x)\rangle_{\alpha}
 \langle \widetilde{\cal O}_{R}(x){\cal O}_3(x_3){\cal O}_4(x_4)\rangle_{\beta}.
\label{eq:BF-CPW-as-intertwiner-gluing}
\eeq
This is precisely the single-valued conformal partial wave \cite{FerraraShadow,SimmonsDuffinShadows,KarateevHarmonic}, with the external representation labels suppressed on the left-hand side.  Thus, once the principal-series Plancherel resolution has been chosen for a spherical cut, its position-space basis is the conformal partial-wave basis in CFT.  The conformal Ward identities, the three-point structures, and the shadow integral are the position-space realization of the same singlet conditions, intertwiners, and invariant contractions that already appeared in the flat BF Wilson-tree calculation.

It follows that changing the conformal partial-wave channel must be represented by the same Racah transformation.  This is indeed what is found in conformal harmonic analysis: the crossing kernel between different partial-wave bases is the conformal $6j$ symbol \cite{GaddeConformalTheories,LiuConformal6j}.  For the scalar, or spherical, subset of this principal series, Krasnov and Louko computed the corresponding Racah coefficients explicitly for $SO_0(1,d+1)$ \cite{KrasnovLoukoRacah}. More general spinning conformal $6j$ symbols are naturally defined by the same intertwiner recoupling problem.  Bissi and Sarkar explicitly derive the Biedenharn-Elliott (pentagon) integral identity obeyed by these conformal Racah kernels \cite{BissiSarkar6j}, realizing in the principal-series harmonic basis the associativity expected abstractly from representation-theoretic recoupling.  The known CFT harmonic-analysis results therefore have exactly the structure predicted by the elementary BF calculation: the conformal partial waves are a position-space realization of the BF channel basis, the conformal $6j$ is its change-of-presentation kernel, and the pentagon identity is the corresponding coherence condition.

Of course, all of the discussion above is kinematic and is naturally formulated on the Euclidean principal-series contour.  The BF principal-series continuum is not the physical CFT spectrum.  But an individual conformal partial wave may be analytically continued away from the principal-series contour\footnote{For short multiplets like conserved currents and the stress tensor, there are some extra subtleties. But that it can be done is all that will matter for us.}, after which the physical lowest-weight/OPE prescription described in the next subsection selects the branch associated with the primary. A related but distinct point is that if one starts from a complete principal-series harmonic expansion of a CFT amplitude and wishes to recover its physical OPE expansion, contour deformations and residues of crossed poles may be required \cite{KarateevHarmonic,CaronHuotInversion}\footnote{There is a useful Ponsot-Teschner analogue of this analytic continuation and contour structure.  The structural analogue of the Euclidean principal-series contour is the standard Liouville/Teichm\"uller line $\alpha=Q/2+iP$, with $P\in\mathbb R$.  When representation labels are continued away from this line -- as occurs, for example, for sub-threshold conical-defect-like states in AdS$_3$/CFT$_2$ -- the Ponsot-Teschner kernel must likewise be analytically continued.  When poles cross the integration contour, the continuation acquires discrete residue contributions: fusion involving degenerate primaries provides a limiting example in which the integral reduces to a discrete sum.}.

Thus the spherical BF calculation fixes the kinematic principal-series partial-wave basis and its conformal $6j$ recoupling kernel.  The physical spectrum and OPE coefficients remain CFT data. The passage from an individual BF partial wave to an ordinary conformal block is provided by the physical boundary prescription, to which we now turn.

\subsection{Boundary Conformal Blocks from BF TQFT}\label{BRS}

The preceding subsection identified the BF sewing basis on a spherical cut: its principal-series resolution gives shadow-symmetric conformal partial waves in position space. But for AdS/CFT, we also need boundary data. A Wilson line that genuinely ends at the physical conformal boundary must terminate in a local conformal primary, and the resulting boundary correlator must obey the OPE asymptotics dictated by that primary. We now formulate this prescription and relate it to the ordinary conformal block.

There is a useful precedent for the connection with BF theory, in the gravitational open-Wilson-network construction of conformal blocks, by Bhatta, Raman and Suryanarayana (BRS) \cite{BhattaWilsonNetworks,BhattaScalarNetworks}.  In the first-order Hilbert-Palatini description used there, the vielbein and spin connection are combined into an $SO(d+1,1)$ connection,
\bea
A=\frac12\omega^{ab}M_{ab}+\frac1\ell e^a M_{\bullet a}\,.
\eea
Here $\ell$ is the physical AdS radius. One begins with a trivalent graph whose edges are labelled by representations of $SO(d+1,1)$. The representations used for the physical CFT families are the non-unitary Euclidean continuations of positive-energy $SO(d,2)$ conformal modules \cite{BhattaScalarNetworks}.  The external edges carry the representations associated with the inserted CFT primaries, while an internal edge carries the representation $(\Delta,J,\ldots)$ of the conformal family exchanged in that channel. At a trivalent junction one uses the conformal three-point intertwiner appropriate to these physical representation labels.\footnote{This is the direct lowest-weight conformal-block construction, distinct from the principal-series BF sewing resolution of the previous subsection. Their relation is described below through analytic continuation and the OPE/monodromy prescription.} The full Wilson network (with a specific prescription for the external legs that we will discuss below), is then evaluated on pure Euclidean AdS$_{d+1}$.  For this background the AdS connection is flat, $F[A]=0$, so locally $A$ is pure gauge.  A Wilson line carried by a representation $R$ therefore reduces to the representation matrix of the corresponding group element, as discussed in \eqref{Wilson-Rep}. The intertwining property cancels the dependence on the internal trivalent vertex positions. Taking the external endpoints to the asymptotic boundary turns the open Wilson network into a function of the boundary insertion points.  BRS show that this boundary function transforms as the corresponding CFT correlator contribution and satisfies the conformal Casimir equation for the conformal family carried by the internal Wilson line.  With their prescriptions, the resulting solution has the OPE behavior of the ordinary global conformal block \cite{DolanOsbornFourPoint,DolanOsbornPartialWaves}.  Thus, although BRS often use the term ``conformal partial wave,'' the object obtained is the conformal {\em block}, rather than the shadow-symmetric Euclidean partial wave discussed in our previous subsection. This distinction is already encoded in their internal-edge contraction: they sum a complete descendant basis in the physical conformal module, with the radial-quantization pairing, rather than gluing two principal-series intertwiners with the shadow pairing. The construction was worked out explicitly for global blocks in $d=2$ in \cite{BhattaWilsonNetworks}. More directly relevant for us, \cite{BhattaScalarNetworks} implements the same idea for four scalar external operators with scalar exchange in general dimension and reproduces the standard scalar conformal block.

The higher-dimensional result is particularly suggestive here.  For $d=2$, vacuum three-dimensional Einstein gravity with negative cosmological constant is locally AdS$_3$, and its first-order formulation is itself a flat-connection Chern-Simons theory.  In higher bulk dimension this is no longer true: the condition $F[A]=0$ is stronger than the Einstein equations and selects locally AdS configurations\footnote{ Section~\ref{sec:asymptoticads} will make the relation between the BF connection and the AdS gravitational sector more explicit.}.  The calculation of \cite{BhattaScalarNetworks}, however, is performed precisely on the pure-AdS background, where the connection is flat.  At the level of this fixed-background Wilson-network calculation, the ingredients are therefore the same as in the Wilson-line sector of our BF theory. Thus the BRS network, with its physical representation labels and endpoint and intertwiner prescriptions, can be evaluated directly on the same flat connection in BF theory.\footnote{This is a statement about the Wilson network, not the sewing resolutions used to construct amplitudes. The spherical BF cut uses the principal-series Plancherel resolution and its shadow pairing, whereas the direct conformal-block construction resolves the intermediate physical conformal family with its descendant Gram pairing. We will explain their connection below.}

We will therefore emphasize that evaluating the network on the pure-AdS flat connection need not itself be regarded as a {\em semiclassical} approximation. What differs is the theory in which this object is embedded. The gravitational construction obtains the flat connection by restricting the Palatini variables to the locally AdS background, whereas in BF theory flatness is a fundamental bulk equation of motion. 

Let us clarify why {\em this} construction of conformal blocks from the bulk, is special for our purposes, compared to some others that are available in the market.  In general dimension, conformal blocks can be represented by geodesic Witten diagrams, in which the interaction vertices of an AdS exchange diagram are restricted to appropriate boundary-anchored bulk geodesics \cite{HijanoGeodesicWitten}.  This is naturally a metric (i.e., effective) description rather than directly a Wilson-network observable\footnote{Even though, in light of our results here, it is tempting to entertain the possibility that geodesics are more primitive than semi-classical spacetime itself, at least in some situations.}. Conversely, Wilson-line constructions of conformal blocks are particularly well developed in AdS$_3$/CFT$_2$, where three-dimensional gravity admits a Chern-Simons formulation and the gravitational variables are themselves encoded in flat connections \cite{BeskenInteractingWilson}. See also \cite{Kostya1,Kostya2} for AdS$_2$ results. The relevance of \cite{BhattaWilsonNetworks,BhattaScalarNetworks} for the present discussion is that they formulate the same conformal representation-theoretic problem directly in terms of higher dimensional open Wilson networks. This makes it plausible that it is a basic bulk object in the full quantum gravity. As we will argue later in this paper, non-trivial local curvature is an effective phenomenon, while the BF conformal block is a kind of ``atom'' of the bulk. 

A key technical point for us will be that the external legs are projected onto appropriate ``cap states'' in the BRS construction. Algebraically, let $G=SO(d+1,1)$ and $K=SO(d+1)$, so that the bulk AdS is the homogeneous space $G/K$. The cap specifies how the free representation index at the open end of an external Wilson line is projected, and provides the endpoint prescription for that leg of the Wilson-network matrix element. It is chosen to transform in a representation of the stabilizer $K$: for a scalar, it is a $K$-singlet. This $K$-equivariance is the natural thing to demand when you want a local bulk field interpretation \cite{BhattaScalarNetworks}. In particular, the scalar cap obeys the $SO(d+1)$ singlet conditions generated by $M_{ij}$ and $P_i-K_i$ \footnote{We use signs adapted to our conformal basis, used elsewhere in the paper.}. The BRS prescription for the external leg \cite{BhattaScalarNetworks} is that the states in the external leg primary module are propagated with the Poincar\'e-patch representative of the flat connection, $g(\rho,x)=e^{-\rho D}e^{-x\cdot P}$, and projected onto the cap state. For the external leg primary, this reduces essentially to
\beq
 \langle\!\langle\Delta|\,g(\rho,x)\,|\Delta\rangle
 =
 e^{\rho\Delta}\bigl(1+e^{2\rho}x^2\bigr)^{-\Delta}
 =
 \left(\frac{z}{z^2+x^2}\right)^\Delta,\qquad z=e^{-\rho}.
\label{eq:BRS-Poisson-kernel}
\eeq
Mathematically \cite{Helgason}, this is the scalar Poisson kernel on $G/K$, which is famously the bulk-to-boundary propagator in semi-classical AdS/CFT \cite{WittenAdS}. The full conformal block is then obtained by assembling these external-leg matrix elements with the trivalent intertwiners and the contractions over the exchanged conformal modules, taking all external endpoints to the asymptotic boundary, and extracting the leading boundary component \cite{BhattaWilsonNetworks,BhattaScalarNetworks}.  For scalar external legs, this is (up to normalization choices) the familiar extrapolate-type limit, and it gives the corresponding conformal block.

For our BF construction, we will replace the cap step by an intrinsic endpoint prescription for the Wilson line, at the conformal boundary. In the semi-classical regime, the local bulk field, the bulk-to-boundary propagator, and the extrapolate dictionary give us the same answer, but those interpretations are not required to obtain the conformal representation-theoretic boundary limit, is our point\footnote{Our proposal is simple, but it is still a {\em proposal}. You cannot get a boundary without defining it. This can be viewed as a UV-complete piece of the bulk/boundary correspondence, that reproduces the semi-classical AdS/CFT result in the appropriate regime.}.  Below we formulate the corresponding endpoint condition intrinsically in BF language.

\subsubsection{Endpoints for Wilson Lines} \label{cuts}

At a physical boundary, unlike a cut, the line genuinely terminates. On the non-degenerate\footnote{This is the analogue of working with an invertible triad in $d=2$ in the gravitational sector.} conformal-Cartan branch of Section \ref{sec:asymptoticads}, with its boundary parabolic reduction, the soldering component of the asymptotic BF connection defines a boundary conformal geometry. Since the BF equation imposes flatness, this Cartan geometry is locally conformally flat (as we will see later) and hence locally modeled on the standard conformal sphere
\bea
 S^d\simeq G/P_{\rm conf}\,,\qquad G=SO_0(d+1,1)\,,
\eea
where $P_{\rm conf}$ is the parabolic subgroup stabilizing a point, equivalently a null ray in the embedding-space description \cite{CapSlovakParabolic,CapSlovakWeyl}. Thus a boundary point $x$ has its usual intrinsic conformal meaning, and the boundary action of $G$ is the standard conformal action.

On this conformal boundary there is a natural endpoint condition: a Wilson line carrying physical conformal representation data $(\Delta,J,\ldots)$ may end at a point $x$ as a local primary ${\cal O}_{R}(x)$ with the same conformal quantum numbers \cite{KarateevHarmonic}. This gives an intrinsic boundary prescription for terminating the Wilson line. Reversing its orientation replaces $R$ by the dual representation (which in radial quantization is the conjugate). Note that a Wilson line corresponding to a physical CFT insertion on the boundary carries the physical conformal-family label $(\Delta,J,\ldots)$, rather than a principal-series label introduced as part of our BF harmonic resolution. Algebraically, the corresponding family is the positive-energy lowest-weight conformal module ${\cal V}_{\Delta,J}$ of the Lorentzian real form $SO(d,2)$, used in the Euclidean Wilson-network calculus by analytic continuation. 

Now consider an open Wilson network with several such endpoints. Each external line ends on a primary ${\cal O}_{R_i}(x_i)$, while the BF junctions use the induced-representation intertwiners of the preceding subsection, analytically continued in the representation labels as needed.  The primary endpoint condition fixes the external label, but by itself it does not distinguish an ordinary conformal block from the shadow-symmetric Euclidean partial wave. The additional datum is a boundary lowest-weight/OPE prescription: in the relevant boundary OPE limit, one selects the branch associated with an exchanged primary of quantum numbers $(\Delta,J)$, rather than its shadow. This choice is part of our boundary prescription\footnote{Clearly, this {\em feels} like an analogue of picking the normalizable mode. But from the bulk Wilson line perspective, there is something a little non-local about this: we are demanding that when two boundary labels approach each other, they should be on the OPE (local) branch.}.

For a spherical cut, the BF sewing resolution is the Euclidean principal-series. Gluing the two three-point intertwiners with the invariant shadow pairing gives the single-valued conformal partial wave $\Psi_{\Delta,J}$ of \eqref{eq:BF-CPW-as-intertwiner-gluing}. Regarded meromorphically in $\Delta$, this partial wave contains the two Casimir branches associated with $(\Delta,J)$ and $(d-\Delta,J^R)$. In reflection-invariant spin sectors, $J^R\simeq J$, so schematically these are $G_{\Delta,J}$ and $G_{d-\Delta,J}$ \cite{SimmonsDuffinShadows,KarateevHarmonic}.  At the physical conformal boundary we impose the OPE condition appropriate to a primary of dimension $\Delta$ and spin $J$: in the OPE limit one keeps the $\Delta$ branch and discards the shadow branch.  In the shadow formalism this is the so-called monodromy projection on the individual partial wave \cite{SimmonsDuffinShadows},
\beq
 {\cal P}_{\Delta}\,\Psi_{\Delta,J}
 \propto G_{\Delta,J}\, .
\label{eq:BF-partial-wave-to-block}
\eeq
Thus, at the level of an individual analytically continued BF partial wave, the boundary OPE/monodromy prescription selects the ordinary conformal-block branch. 

But now, after this analytic continuation and OPE branch selection, the same block can be written directly in radial quantization \cite{SimmonsDuffinShadows,CaronHuotInversion}. The primary of quantum numbers $(\Delta,J)$ generates the positive-energy lowest-weight conformal module ${\cal V}_{\Delta,J}$. After quotienting null states, let $G_{ab}$ denote its non-degenerate descendant Gram matrix.  The projector onto the conformal family is schematically,
\beq
 \Pi_{\Delta,J}
 =
 \sum_{a,b\in{\cal V}_{\Delta,J}}
 |\Delta,J;a\rangle\,(G^{-1})^{ab}\,
 \langle\Delta,J;b|\, .
\label{eq:physical-family-projector}
\eeq
Contracting the corresponding kinematic three-point matrix elements through
$\Pi_{\Delta,J}$ gives $G_{\Delta,J}$.  This is the form used in the BRS Wilson-network construction, where the descendant indices of the exchanged conformal family are contracted with the module pairing
\cite{BhattaWilsonNetworks,BhattaScalarNetworks}.  It is the radial-quantization expression for the same $\Delta$ branch selected in \eqref{eq:BF-partial-wave-to-block}.

Our BF/CFT proposal is therefore to keep the principal-series BF resolution as the gluing resolution of a spherical cut. At the level of an individual analytically continued partial wave, the boundary OPE/monodromy prescription selects the physical conformal-block branch. For a complete amplitude, the physical OPE expansion is recovered from the principal-series harmonic expansion by meromorphic continuation of its coefficient density and the corresponding contour deformation, including various complications like residues or isolated contributions as required. For a selected physical family, the block may be evaluated directly with the lowest-weight projector \eqref{eq:physical-family-projector}. We expect that a similar approach can work  in the BF resolutions associated with the other admissible cuts and presentations in the sewing-closed class $\mathfrak C$. 

The BRS cap projection and extrapolate limit, fit naturally into this as a bulk-EFT-motivated prescription. Fundamentally, no EFT field equation or extrapolate dictionary need be involved in the underlying representation-theoretic statement. BF-TQFT simply gives a bulk mechanism for organizing this representation theory: but in the semi-classical regime, one may re-interpret it via a local bulk field or cap, a bulk-to-boundary wavefunction, and the usual extrapolate limit.

\subsection{Further Evidence: Higher Topology, Holonomy and Defects}\label{section:higher}

The evidence in the previous subsections was concrete, but relatively ``local". The remaining question is more global. Can the corresponding BF gluing calculus continue to make sense when a CFT cut has non-trivial topology? It turns out that there are several BF precedents which address these ingredients in various contexts. Even though not specifically for BF theory with our particular gauge group, we find the evidence to be extremely suggestive\footnote{Let us emphasize one technical point before we proceed with this subsection: A non-contractible cycle can carry global flat-connection holonomy without any magnetic defect being present. A codimension-two defect is a prescription for a local meridional monodromy/holonomy around its worldvolume. They nevertheless belong naturally to the same sewing calculus: as the examples below illustrate, the same boundary cut can admit complementary fillings in which a boundary cycle is respectively a genuine non-contractible cycle or a contractible meridian linking a defect. Loosely speaking, codim-2 defects introduce sources in the $A$-flatness equation, while Wilson lines introduce sources in the $B$-equation.}.

A key useful bridge between local recoupling and global sewing is a \emph{state sum}. Conceptually, a state sum implements repeated BF gluing by decomposing the manifold into elementary pieces. After choosing a triangulation or ``cell decomposition'', one labels its internal cells by representation and intertwiner data, multiplies the corresponding local simplex amplitudes, and sums or integrates over the internal labels\footnote{Choosing how representation and intertwiner data are assigned to the cells of the decomposition, together with a prescription for contracting and summing these data when the elementary pieces are glued, so that changes of triangulation are realized by recoupling identities, is the non-trivial part of the construction.}. We will not be too explicit about this because we only need the results. But the picture is intuitive and the conclusion is relevant: in a topological BF state sum the triangulation itself is auxiliary. Local recoupling and completeness identities imply that different triangulations give equivalent evaluations. The state sum thus provides a concrete mechanism by which local recoupling coherence propagates through repeated gluings, so that the resulting amplitude is determined by its topology (and other global data). In other words, there is good reason to believe, that local coherence is sufficient to assign consistent BF amplitudes to manifolds with non-trivial topology.

The most relevant higher-dimensional example for us is Ooguri's four-dimensional $SU(2)$ BF model \cite{Ooguri4dBF}. Its amplitudes are obtained by contracting representation and intertwiner data over a triangulated four-manifold, with changes of channel governed by the corresponding (underlying) $6j$ recoupling coefficients. After the BF gauge volume divergences are accounted for, the remaining quantity depends only on the piecewise-linear topology. Thus the local recoupling identities are strong enough to make a large collection of internal channel sums independent of how the full four-manifold was triangulated.

Ooguri also constructs states on the three-manifold
$Y=S^2\times S^1$ that are directly relevant to our non-spherical cuts\footnote{A complementary three-dimensional precedent appears in Ooguri's analysis of Ponzano-Regge gravity, where general cuts appear \cite{Ooguri3dTopology}. But since this is morally somewhat analogous to the Virasoro TQFT case, we will not consider it as genuine evidence for non-spherical cuts in the $d>2$ case.}. In one construction the boundary is filled by $D^3\times S^1$ and a Wilson loop winds the non-contractible $S^1$.  Since $\pi_1(S^2\times S^1)=\mathbb Z$, a flat $SU(2)$ connection may carry a conjugacy class $[g]$ of holonomy around this cycle, and a spin-$j$ Wilson loop probes it through the character $\chi_j(g)$.  Ooguri also considers the complementary filling $S^2\times D^2$.  Here the boundary $S^1$ is contractible in the filling, and a non-trivial meridional holonomy is instead supported by a 't Hooft surface on the core $S^2$. Both constructions produce BF states on $S^2\times S^1$ that Ooguri calls \emph{physical states}, namely states whose insertion does not spoil the topological invariance of the four-dimensional state sum \cite{Ooguri4dBF}. They therefore give a concrete higher-dimensional precedent for a single non-spherical BF state space admitting both electric and magnetic preparations: one by a Wilson loop probing the global holonomy and the other by a magnetic defect supporting the defect monodromy.

Let us also note something that Ooguri would have found underwhelming in '92: the same boundary topology, $S^2\times S^1$, is the Euclidean boundary topology of thermal AdS$_4$. In the conformal $SO(4,1)$ problem, the thermal Cartan holonomy is fixed by the boundary conformal data. If this $S^1$ remains non-contractible, the prescribed boundary holonomy is realized directly as the holonomy of the flat connection around that cycle. In a complementary filling where the circle bounds a disk, we have a codimension-two core defect. We believe this is the BF analogue of the Hawking-Page cycle-switching in black holes, but with a twist: defects are likely to play an important role.

In fact, in the compact-group setting the relation between these two preparations can be made explicit by Peter-Weyl harmonic analysis, which we interpret as two ways of doing BF quantization.  Let $\Psi_{[h]}(g)$ denote the class-function state (with $g$ a boundary holonomy variable) in the $S^2\times D^2$ filling which fixes the meridional holonomy to the conjugacy class $[h]$.  With normalized Haar measure, the corresponding conjugacy-class delta function has the character expansion
\bea
\Psi_{[h]}(g)=\delta_{\rm cl}(g,[h])
=\sum_{R\in\widehat G}\overline{\chi_R(h)}\,\chi_R(g)\,.\label{Peter-Weyl}
\eea
The functions $\chi_R(g)$ are the wavefunctions obtained in the complementary $D^3\times S^1$ filling from Wilson lines in irreducible representations $R$ wrapping the non-contractible $S^1$.  Thus the two preparations are related by a universal BF change-of-presentation kernel $K([h],R)=\overline{\chi_R(h)}$, independent of any boundary CFT data.  This gives a simple explicit compact-group model for the non-spherical changes of channel that we have assumed. A key remaining step is to interpret (conformal) BF class-functions or their generalizations as CFT objects associated to a suitable class of basis states in the dual channel. We suspect that an appropriate non-compact analogue of this transform may lead to an understanding of Cardy-like physics for the black hole density of states in higher dimensions \cite{Panikkar}.  

Changing gears: Cattaneo, Mnev and Reshetikhin (CMR) give general arguments that BF theory can be consistently cut and glued even on manifolds with non-trivial topology and flat-connection holonomies \cite{CattaneoMnevReshetikhinCellular}.  Their general non-Abelian construction is perturbative around a chosen flat connection, so it does not provide a full non-perturbative quantization of the holonomy moduli, but it shows explicitly that such global degrees of freedom are compatible with the sewing structure.

Non-compactness of the gauge group is a separate issue, and there are useful precedents there as well.  Freidel's Lorentzian Ponzano-Regge model realizes $SO(2,1)$ BF theory as a triangulation-independent state sum involving both discrete and continuous representation sectors and the corresponding $SL(2,\mathbb R)$ recoupling coefficients \cite{FreidelLorentzianPR}. In the canonical description the transition amplitude is an integral over the moduli space of flat connections on the spatial surface, while the same Lorentzian BF theory admits a non-compact Plancherel decomposition. Thus non-compact harmonic analysis is compatible with the quantization of global flat-connection degrees of freedom. More generally, gauge-invariant spin-network Hilbert spaces and natural measures have been constructed for non-compact groups including $SL(2,\mathbb R)$ and $SL(2,\mathbb C)$ \cite{FreidelLivineNoncompact}.  These results address some of the analytic difficulties caused by continuous representation labels and non-compact gauge orbits.

Finally, higher-categorical state sums give non-BF examples in which local higher-categorical data assemble into a four-dimensional state sum invariant under changes of triangulation \cite{DouglasReutter}. They reinforce the general local-to-global lesson of the BF examples above. Taking stock: compact four-dimensional BF shows global recoupling together with the possibility of an explicit character transform between Wilson and defect preparations on the same non-spherical cut, CMR shows cobordism/gluing mechanisms in the presence of flat-bundle data, and three-dimensional Lorentzian BF shows that non-compact harmonic analysis can be incorporated into a global state sum. These results provide evidence that global flat-connection degrees of freedom and defect-decorated state spaces can be incorporated in a single sewing-consistent BF calculus. 

There is also evidence from CFT that consistency conditions on non-spherical backgrounds are closely connected to the familiar spherical conformal-block structure. We will not get into the details here, but will mention some papers which address closely related questions in the context of thermal partition functions and correlators \cite{ShaghoulianGeneralizedCardy,HorowitzShaghoulianDetachable,BuricThermal,AlkalaevMandryginThermal,AlkalaevMandryginSamsonov}. 

\subsection{The BF Gluing Bootstrap}
\label{subsec:BF-bootstrap-full}

Let us now formulate the bootstrap. Let
\beq
\mathfrak X=(X^d,\mathfrak d_\partial)
\label{eq:BF-boundary-datum-bootstrap}
\eeq
be an admitted asymptotic CFT configuration. Each presentation $\mathfrak p$ of $\mathfrak X$ defines a BF block basis $\{\Psi^{\rm BF}_{\mathfrak p,\lambda}\}$, as described above. For a CFT with physical data ${\mathsf D}$, its amplitude in an admissible BF presentation $\mathfrak p$ has a generalized coefficient distribution
\beq
 \rho_{\mathfrak p}[\mathsf D](\lambda),
\label{eq:BF-coefficient-distribution}
\eeq
relative to the BF block resolution, so that\footnote{The integral is understood in a generalized sense: the BF resolution may include discrete or isolated contributions in addition to continuous pieces, together with the appropriate contour/residue prescription after analytic continuation.}
\beq
{\cal A}_{\mathsf D}(\mathfrak X;\mathfrak p)
 =
 \int d\mu_{\mathfrak p}(\lambda)\,
 \rho_{\mathfrak p}[\mathsf D](\lambda)\,
 \Psi^{\rm BF}_{\mathfrak p,\lambda} .
\label{eq:BF-channel-amplitude-bootstrap}
\eeq
Here $\rho_{\mathfrak p}$ is simply the coefficient distribution of the CFT amplitude in the chosen BF resolution. When $\mathfrak p$ has an independent physical channel interpretation, this distribution is fixed by the corresponding physical CFT data. In particular, in a spherical local channel it is determined by the usual spectrum and OPE data: the same correlator may be expressed in the principal-series BF harmonic basis by the corresponding harmonic decomposition.  For a generic admitted BF presentation no separate physical spectrum may be known, $\rho_{\mathfrak p}$ is then simply the representation of the same amplitude in that BF basis.

The BF blocks and coefficient distributions are presentation-dependent, while their contraction in \eqref{eq:BF-channel-amplitude-bootstrap} represents the fixed CFT amplitude. The bootstrap condition is therefore presentation independence within the admitted BF calculus:
\beq
\begin{gathered}
 {\cal A}_{\mathsf D}(\mathfrak X;\mathfrak p)
 =
 {\cal A}_{\mathsf D}(\mathfrak X;\mathfrak p')\,,\\
 \text{for every admitted }\mathfrak X
 \text{ and every pair of admissible presentations }\mathfrak p,\mathfrak p'\text{ of }\mathfrak X
 \end{gathered}
\label{eq:BF-master-sewing-bootstrap}
\eeq
Whenever the BF kernel admits an invertible continuation to the generalized coefficient distributions and contour prescriptions appearing above, this presentation-independence condition can be represented at the level of BF coefficients as
\beq
\rho_{\mathfrak p'}[\mathsf D](\lambda')
 =
 \int d\mu_{\mathfrak p}(\lambda)\,
\rho_{\mathfrak p}[\mathsf D](\lambda)\,
 K^{\rm BF}_{\mathfrak p\to\mathfrak p'}
 (\lambda,\lambda';\mathfrak X),
\label{eq:BF-density-sewing}
\eeq
with the integral understood in the same analytically continued or distributional sense as in \eqref{eq:BF-channel-amplitude-bootstrap}. The kernel is an in-principle known object in the calculus, for the sphere cut bootstrap it is the Racah-Wigner 6j symbol we discussed earlier. For a generic target presentation $\mathfrak p'$, \eqref{eq:BF-density-sewing} simply transports the coefficient distribution into another BF resolution. Its bootstrap content arises when $\mathfrak p'$ has an independently specified physical channel expansion, or more generally when the same physical CFT data recur in different presentations: the transported distribution must then agree with the one dictated there by the same data $\mathsf D$. The bootstrap problem is to find $\mathsf D$ satisfying all such compatibility conditions. In the four-point spherical Wilson-line subsector this is ordinary global conformal crossing.

In ordinary CFT, it is expected that the amplitudes and state spaces do exist on the admitted backgrounds -- so we do not view {\em that}, as being a part of {\em our} conjecture, even though explicit spectrum data are generally not known. Our proposal is that their kinematic changes of presentation are organized by one universal, defect-closed conformal BF calculus. This is realized directly in the spherical Wilson-line sector by conformal harmonic analysis, conformal $6j$ kernels and Wilson-network constructions, and is supported structurally at higher topology by the observations we presented earlier.

For it to define a conventional local CFT, a bootstrap solution must also obey other usual physical requirements: locality and OPE convergence where a local OPE exists, a non-degenerate pairing, reflection positivity or unitarity as appropriate, the identity operator, and a conserved stress tensor satisfying its Ward identities \cite{PappadopuloOPE}. The stress-tensor normalization $C_T$ and the physical spectra and matrix elements are part of the solution space, they are not fixed by the BF kinematics. Large-$C_T$, sparse-spectrum or semiclassical conditions relevant to holographic gravity are further restrictions on the space of solutions. We will be only interested in theories with a stress tensor in their spectrum.

\subsubsection{When Does Defect Bootstrap Kick In?}
\label{whenboot}

We expect that defect-BF quantized objects are a natural way to describe the physical spectrum of a CFT in non-spherical channels.

The familiar torus bootstrap in $d=2$ is useful for clarifying both the analogy and the difference.  Schematically, one writes
\beq
 Z(\tau,\bar\tau)=\sum_{i\bar j} n_{i\bar j}\,\chi_i(\tau)\,\bar\chi_{\bar j}(\bar\tau),
 \qquad
 \chi_i(-1/\tau)=\sum_j S_{ij}\chi_j(\tau),
\eeq
and modular invariance requires $Z(\tau,\bar\tau)=Z(-1/\tau,-1/\bar\tau)$.  This is a bootstrap condition because the dual quantization is independently required to be another description of the same CFT, with the {\em same} physical spectrum reorganized by the modular kernel.  In $d=3$, for example on $S^1\times S^2$, there is no analogous modular symmetry that by itself identifies the physical spectra in the two presentations.  Nevertheless the BF sewing problem can have the same logical structure. It is natural (see Section \ref{sec:tower}) to assign a candidate local/thermal spectrum to $D^3\times S^1$, while the complementary $S^2\times D^2$ presentation with a core monodromy defect may be assigned an independently admissible defect spectrum.  Neither spectrum is known a priori: both are part of the CFT data to be bootstrapped\footnote{Note that from the bulk, the CFT spectrum in the local channel is {\em also} a postulated statement about the spectrum of defects -- namely Wilson lines. But we are comfortable with it, because from the boundary, we understand them better. They are local operators.}.  If {\em both} presentations have such physical interpretations, the universal BF kernel relating the presentations will lead to a bootstrap condition, when the amplitudes are equated. Note that only after an independent claim about the spectrum of states in the dual channel, does it actually become a bootstrap condition. The resulting constraint then relates two a priori distinct sets of physical data, rather than closing on a single character spectrum as in the modular case.  In this sense modular invariance is a particularly powerful realization of the more general principle: bootstrap begins when distinct BF presentations carry independently specified physical channel data whose amplitudes must agree.

\section{The AdS Gravitational Sector}
\label{sec:asymptoticads}

Section~\ref{sec:BF-bootstrap} treated conformal BF theory as the kinematic sewing skeleton underlying the bootstrap.  We now explain how an AdS gravitational interpretation emerges from the same variables.  The essential extra choice is the so-called conformal-Cartan branch of the connection: at the boundary we choose the parabolic reduction appropriate to conformal geometry, and require the resulting soldering component of the connection to be non-degenerate.  It then defines a conformal geometry.  For $d>2$ there is no further need for a gravitational or Drinfel'd-Sokolov reduction: $F$-flatness on this branch already implies the {\em normal} conformal-Cartan conditions and restricts the induced conformal structure to be locally conformally flat. 

We then use this geometric interpretation to relate Wilson lines to quantum states and BF fluxes to their charges. Flat-connection holonomies around non-trivial (e.g., thermal) cycles are also natural physical data.

\subsection[The Non-Degenerate Conformal-Cartan Branch]{The Non-Degenerate Conformal-Cartan Branch}
\label{subsec:conformal-radial}

We now make the geometric branch of the conformal BF connection explicit near the asymptotic boundary.  We choose a boundary-adapted basis for the gauge algebra and a radial gauge, select the branch with non-degenerate soldering form, and then use the BF flatness equation to determine the remaining Cartan data. We will see that much of the story below has analogues in AdS$_3$ even though there is no Brown-Henneaux/DS reduction.

Split the Euclidean internal index as $a=(i,\hat r)$, with $i=1,\ldots,d$, and let $\bullet$ denote the distinguished timelike internal direction.  Introduce the conformal basis
\[
M_{ij}\defi J_{ij}\,,\qquad
D\defi J_{\hat r\bullet}\,,\qquad
P_i\defi J_{i\bullet}+J_{i\hat r}\,,\qquad
K_i\defi J_{i\bullet}-J_{i\hat r}\,.
\]
With our conventions\footnote{Our sign conventions are those of parabolic geometry, which has some unfortunate differences from the physics literature.},
\beq
[D,P_i]=-P_i\,,\qquad
[D,K_i]=+K_i\,,\qquad
[M_{ij},D]=0\,,
\label{eq:conf-algebra-1}
\eeq
\beq
[M_{ij},P_k]=\delta_{jk}P_i-\delta_{ik}P_j\,,\qquad
[M_{ij},K_k]=\delta_{jk}K_i-\delta_{ik}K_j\,,
\label{eq:conf-algebra-2}
\eeq
\beq
[P_i,K_j]=2(M_{ij}-\delta_{ij}D)\,.
\label{eq:conf-algebra-3}
\eeq
Thus
\beq
\mathfrak{so}(d+1,1)
=\mathfrak g_{-1}\oplus\mathfrak g_0\oplus\mathfrak g_{+1}\,,\qquad
\mathfrak g_{-1}=\operatorname{span}\{P_i\}\,,\quad
\mathfrak g_0=\operatorname{span}\{M_{ij},D\}\,,\quad
\mathfrak g_{+1}=\operatorname{span}\{K_i\}\,.
\label{eq:conformal-grading}
\eeq
This is the $|1|$-grading of conformal Cartan geometry, and the parabolic subalgebra $\mathfrak g_0\oplus\mathfrak g_{+1}$ stabilizes a null ray \cite{CapSlovakParabolic,CapSlovakWeyl}. Together with the non-degeneracy of the $P_i$ component below, this boundary-adapted structure is what we mean by the conformal-Cartan branch.

Locally, we can choose the gauge
\[
A=b^{-1}\bigl(a(\rho,x)+d\bigr)b\,,\qquad
b(\rho)=e^{(\rho/\ell)D}\,,\qquad
a_\rho=0\,.
\]
This fixes $A_\rho=D/\ell$, but does not make the tangential connection $a(\rho,x)$ independent of $\rho$.  The grading gives
\[
b^{-1}P_i b=e^{\rho/\ell}P_i\,,\qquad
b^{-1}K_i b=e^{-\rho/\ell}K_i\,.
\]
At the purely conformal-Cartan level, $\ell$ is an arbitrary normalization scale for the $|1|$-graded connection and the radial coordinate.  When we specialize to the AdS gravitational sector, we choose this normalization to coincide with the physical AdS radius used in the BRS discussion earlier and in Appendix~\ref{app:BF}.  At any fixed $\rho$, the tangential connection can be decomposed as
\beq
 a
=\frac1\ell e^iP_i
+\frac12\omega^{ij}M_{ij}
+\phi D
+\frac1\ell f^iK_i\,,
\label{eq:general-a}
\eeq
where the coefficients may still depend on $\rho$:  equation~\eqref{eq:general-a} is merely an algebraic decomposition of an $SO(d+1,1)$ connection.

We now select the geometric branch by requiring $e^i{}_{\mu}$ to be non-degenerate.  It then acts as a soldering form, and after choosing a Weyl frame, defines a metric $g$ and hence a conformal class $[g]$.  This is the higher-dimensional analogue of requiring an invertible dreibein in the Chern-Simons description of AdS$_3$.  At this stage $\omega^{ij}$, $\phi$, and $f^i$ are simply the remaining components of the same Cartan connection. Their geometric form will follow from flatness.

For $d>2$, the remaining local Cartan data are fixed by the $F=0$ flatness equation.  Because $e^i{}_{\mu}$ is invertible, a local $K_i$ gauge transformation can be used to set the dilatation component to zero, $\phi=0$.  The $P_i$ component of $F[a]=0$ is then the torsion-free condition and fixes the Levi-Civita spin connection $\omega^{ij}=\omega_{\rm LC}^{ij}[g]$.  The Ricci contraction of the $M_{ij}$ component fixes $f^i=-(\ell^2/2)\mathsf P^i[g]$, defined below in terms of the Schouten tensor. The $D$ component of $F[a]$ then vanishes because the Schouten tensor is symmetric.  For $d\geq4$ the remaining traceless $M_{ij}$ curvature is the Weyl tensor, which flatness sets to zero.  In $d=3$ the Weyl tensor vanishes identically, and the $K_i$ component of flatness demands the vanishing of Cotton tensor, the conformal-flatness obstruction. (We will not need the terms showing up in $F[a]$, so we will not write them.) The relevant curvature combination for us is the Schouten tensor
\beq
\mathsf P_{ij}[g]
\defi
\frac1{d-2}\left(R_{ij}-\frac{R}{2(d-1)}g_{ij}\right)\,.
\label{eq:Schouten}
\eeq
Writing $\mathsf P^i=\mathsf P^i{}_j e^j$, the connection is therefore
\beq
 a[g]
=\frac1\ell e^iP_i
+\frac12\omega_{\rm LC}^{ij}[g]M_{ij}
-\frac{\ell}{2}\,\mathsf P^i[g]K_i\,.
\label{eq:normal-cartan-flat}
\eeq
Equation~\eqref{eq:normal-cartan-flat} is the form taken by the flat BF connection on the non-degenerate branch in the Weyl gauge $\phi=0$. In fact, equation~\eqref{eq:normal-cartan-flat} is the standard local form, in the Weyl gauge $\phi=0$, of the normal conformal Cartan connection associated with a general conformal class $[g]$ \cite{CapSlovakParabolic,CapSlovakWeyl}. In BF theory, $F$-flatness further restricts $[g]$ to be locally conformally flat.

It remains to use the mixed radial component of the same BF equation.  Since curvature transforms covariantly in radial gauge,
\[
F[A]=b^{-1}F[a]b\,,\qquad
F[a]_{\rho\mu}=\partial_\rho a_\mu\,.
\]
The mixed equation $F_{\rho\mu}[A]=0$ therefore implies $\partial_\rho a=0$: the asymptotic connection is transported unchanged along the radial direction. On a BF saddle we may therefore write
\beq
A=b^{-1}(a+d)b\,,\qquad
b(\rho)=e^{(\rho/\ell)D}\,,\qquad
\partial_\rho a=0\,.
\label{eq:radial-gauge}
\eeq
This is again the same basic mechanism as in the Chern-Simons description of AdS$_3$: radial gauge fixes the radial component, while flatness removes the residual radial evolution.

As we already alluded to, the tangential equations used above have a direct geometric interpretation. Flatness of the normal conformal Cartan connection is precisely local conformal flatness: for $d\geq4$ the Weyl tensor must vanish, while in $d=3$ the obstruction is the Cotton tensor.  Hence, for $d>2$,
\beq
 F[A]=0
 \qquad\Longleftrightarrow\qquad
 [g] \ \text{locally conformally flat on the non-degenerate geometric branch}.
\label{eq:normal-cartan-flatness}
\eeq
Thus the two components of the same BF equation play distinct roles: $F_{\rho\mu}=0$ removes the radial dependence, while $F_{\mu\nu}=0$ becomes the local conformal-flatness condition  on the non-degenerate geometric branch.

Let us make things explicit for the global-AdS cylinder in Euclidean signature
\beq
 ds^2_{\rm cyl}=d\tau^2+R_\partial^2d\Omega_{d-1}^2\,,
\label{eq:cylinder-metric}
\eeq
with Lorentzian continuation $ds^2_{\rm cyl}=-dt^2+R_\partial^2d\Omega_{d-1}^2$.  Writing $e^0=d\tau$ and $e^\alpha=R_\partial\hat e^\alpha$, substitution in~\eqref{eq:normal-cartan-flat} gives
\beq
\begin{aligned}
a_{\rm cyl}={}&
\frac1\ell\left(e^0P_0+e^\alpha P_\alpha\right)
+\frac12\hat\omega^{\alpha\beta}M_{\alpha\beta}
\\[2mm]
&+\frac{\ell}{4R_\partial^2}e^0K_0
-\frac{\ell}{4R_\partial^2}e^\alpha K_\alpha\,.
\end{aligned}
\label{eq:a-cylinder}
\eeq
Here, $\hat \omega^{\alpha\beta}$ is the Levi-Civita spin connection on the round $S^{d-1}$. The cylinder metric is curved but conformally flat, so its normal Cartan connection is locally flat and hence locally pure gauge.  Global holonomies can nevertheless remain around non-contractible cycles or codimension-2 defects.  Section~\ref{sec:tower} will use precisely this longitudinal thermal holonomy.

\subsection{Gravity Eigen-States and the $B$-field Gauss Law}
\label{subsec:wilson-line-center}

Eigenstates of a CFT on a cylinder are organized into unitary lowest-weight conformal modules: each module consists of a lowest-weight primary together with its descendants. We can construct them using Wilson lines. Consider the Euclidean half-cylinder geometry, topologically
\[
M_-\simeq(-\infty,0]_\tau\times B^d\,,\qquad
Y=\{\tau=0\}\simeq B^d\,,
\]
and let a Wilson line $C$ in representation ${\cal R}$ run along the central Euclidean-time direction, meeting the final cut at $p\in Y$.  With the endpoint representation index included among the marked-point data, the cut-open BF path integral schematically prepares
\beq
 |\Psi_{\cal R}\rangle_{\rm BF}
\ \defi\
\int_{M_-}[DA\,DB]\,W_{\cal R}(C)[A]\,e^{-S_{\rm BF}[A,B]}
\ \in\
{\cal H}_{\rm BF}(Y;p,{\cal R})\,.
\label{eq:BF-Wilson-state}
\eeq
A choice of boundary polarization on $Y$ is understood.  This is the higher-dimensional counterpart of the state-preparation construction of \cite{Rajdeep}.  

A key point in the construction of \cite{Rajdeep} is that one starts with unitary Lorentzian representations, keeps them intact while defining the (Euclidean) path integral, but rotates the (time) coordinate. This avoids complications that can arise when trying to define a path integral for the complexified gauge group directly, by identifying a contour in field space \cite{WittenComplex}. We will follow the same idea here. It also means that the labels ${\cal R}$ of our Wilson lines live in the unitary $SO(d,2)$ representations (the lowest-weight modules mentioned above) rather than the unitary principal series ones of $SO(d+1,1)$. Such path integrals naturally produce the primaries and descendants familiar in Lorentzian signature\footnote{More precisely: we mean that the representation space and conformal-generator action are those inherited from the original unitary Lorentzian $SO(d,2)$ module. The resulting Euclidean Wilson-line representation is therefore generally non-unitary, but is obtained by analytic continuation of the same Lorentzian representation data. The cut-open state is interpreted as a Lorentzian state.}.

There is one subtlety compared to the construction in \cite{Rajdeep}. In \cite{Rajdeep} the unitary Lorentzian labels were discrete and continuous series representations of $SL(2,\IR) \times SL(2,\IR)$. Upon quantization via the path integral, this naturally leads to current algebra primaries which turn into the (unique) Virasoro primary after Drinfel'd-Sokolov reduction. A key point is that at the current algebra level, all affine primaries have the same Sugawara weight and therefore the (topological) path integral cannot distinguish them. But the DS reduction removed (or quotiented) those modes, resulting in the (Virasoro) primary. Here, on the other hand, we have no DS reduction: but as we will now argue, none is required.  

Because the Wilson line in the path integral only couples to $A$ and not to $B$, we can integrate out the $B$ and this enforces flatness of the connection. This means that the Wilson line of the cylinder Cartan connection from the previous subsection, is simply the parallel transporter of the state at the other endpoint. Indeed, from~\eqref{eq:a-cylinder}, along the Euclidean-time direction, one has up to conjugation
\bea
A_\tau \sim \frac{D}{R_\partial},
\eea
where $\sim$ denotes conjugacy in the conformal group. Truncating the half-cylinder at $\tau=-T$, parallel transport of the Wilson-line state to $\tau=0$ acts as 
\bea
\exp\!\left(\frac{T}{R_\partial}D\right).
\eea
With our convention $[D,P_i]=-P_i$, a primary in a lowest-weight module obeys $D|\Delta,J\rangle=-\Delta|\Delta,J\rangle$, while its level-$n$ descendants have eigenvalue $-(\Delta+n)$.  After removing the overall factor $e^{ -T\Delta/R_\partial}$, the descendants are therefore suppressed by $e^{-nT/R_\partial}$ as $T\to\infty$.  The semi-infinite Euclidean path integral consequently projects onto the lowest-weight primary subspace (or with the endpoint polarization selecting a component when $J\neq0$). The story is therefore quite parallel to the $d=2$ picture in \cite{Rajdeep}. In $d>2$, boundary gravitons, which should not be present, are indeed not present.

Let us identify where the Wilson charge lives in BF theory. Semiclassically a Wilson line may be described by a coadjoint-orbit variable $u$ and a representative $\Lambda$ \cite{AlekseevFaddeevShatashvili},
\beq
S_{\rm WL}[A;u,\Lambda]
=\int_C\left\langle\Lambda,\,u^{-1}(d+A)u\right\rangle\,,
\qquad
q=u\Lambda u^{-1}\,.
\label{eq:WL-action}
\eeq
Writing $J_C=q\,\delta_C$ for the Poincar\'e-dual current, variation of the BF action with the line inserted gives,
\beq
F[A]=0\,,\qquad
D_A B+\kappa_{\rm BF}J_C=0\,.
\label{eq:sourced-BF}
\eeq
The line therefore does not change the local flatness equation for the conformal-Cartan connection.  It selects a $B$-charge sector. This labels a state rather than a backreacted metric geometry. The latter, we expect is an ensemble state suitable for describing the effective dynamics of light states interacting with a heavy microstate. When a gravitational EFT exists, any geometric backreaction should be described in terms of the EFT. This is related to the presence of the stress tensor (and its multi-trace tower) in the spectrum, and is discussed more in Section~\ref{sec:ward-simplicity}. A full understanding of the origin of non-trivial local curvature and its dynamics from these structures, is clearly a very interesting direction to pursue.

Let $D^d$ be a small transverse ball meeting $C$ once and $S^{d-1}_{\rm link}=\partial D^d$.  For any adjoint valued field that satisfies $D_A\zeta=0$ on the punctured ball, the sourced Gauss law gives 
\beq
\int_{S^{d-1}_{\rm link}}\langle\zeta,B\rangle
=-\kappa_{\rm BF}\int_{D^d}\langle\zeta,J_C\rangle\,.
\label{eq:linking-flux}
\eeq
Thus the Wilson representation is measured locally by the linking $B$-flux.  In a flat trivialization this may be summarized invariantly as
\beq
\left[\int_{S^{d-1}_{\rm link}}B\right]
=-\kappa_{\rm BF}[\Lambda] \, ,
\label{eq:B-flux-orbit}
\eeq
where the brackets retain only the conjugacy/coadjoint-orbit data.

This is one way in which the higher-dimensional story differs from the familiar three-dimensional holonomy intuition.  For $d>2$ the link of a worldline is $S^{d-1}$, which is simply connected, so there is no generic meridional $A$-holonomy associated with the electric line.  The same sphere does support the $(d-1)$-form flux in~\eqref{eq:linking-flux}.  By contrast, the codimension-two magnetic defects of Section~\ref{subsec:BF-electric-magnetic-completion} have an $S^1$ meridian and are naturally specified by $A$-monodromy.  A genuine non-contractible cycle of the spacetime, such as the thermal $S^1_\beta$, may of course also carry global $A$-holonomy.  This separation will be used directly below.  Section~\ref{sec:tower} uses the global thermal holonomy of $A$ to organize Wilson loops into conformal characters and multi-trace towers.  Section~\ref{sec:ward-simplicity}  gives the representation-valued $B$-flux its physical boundary interpretation: its conserved currents are identified with the conformal currents built from the CFT stress tensor.  

\section{Wilson Towers as Bulk Local Fields}
\label{sec:tower}

The BF skeleton described above is topological, but perturbative bulk local fields can emerge in bootstrapped theories, if their spectrum contains towers of multi-trace primaries. The relevant idea was discussed recently in AdS$_3$, by re-interpreting the ``Wilson spool" as a ``Wilson tower". The familiar one-loop determinant of a free field can be represented by a multiply wound Wilson spool in AdS$_3$ \cite{GMY,CastroComanFlissZukowskiCoupling,FlissSpoolQuotients}. That spool can in turn be reorganized into a sum of \emph{single}-winding Wilson loops labelled by the multi-trace primaries of the dual generalized-free field \cite{Vishal}.  We now show that the same mechanism has a direct higher-dimensional realization in the conformal-Cartan variables of Section~\ref{sec:asymptoticads}.

On thermal AdS$_{d+1}$, the one-particle Hilbert space of a free bulk field is a lowest-weight representation of $SO(2,d)$.  The thermal trace of that representation is therefore a conformal character.  But the same character is also the trace of the conformal-Cartan holonomy around the non-contractible thermal circle.  The ordinary bosonic Fock-space identity then turns the one-loop determinant into a plethystic exponential of this Wilson loop, while the cycle-index decomposition reorganizes the result into Wilson loops labelled by multi-trace conformal primaries. The separate ingredients -- thermal-AdS determinants, $SO(2,d)$ character descriptions, and the multi-trace/Fock-space interpretation -- are standard \cite{GopakumarGuptaLalHeatKernel,KakkarSarkarAdS, SunAdSCharacters,FitzpatrickKaplanAdSField}. Here we directly realize them as Wilson observables of the higher-dimensional conformal BF skeleton, and link all these perspectives together.

\subsection{The Thermal Circle as a Conformal-Cartan Wilson Loop}
\label{subsec:thermal-wilson-character}

Consider Euclidean thermal AdS$_{d+1}$,
\beq
M_\beta=\mathbb H^{d+1}/\mathbb Z\,,\qquad
\partial M_\beta=S^1_\beta\times S^{d-1}\,,
\label{eq:thermal-ads-quotient}
\eeq
with the generator of $\mathbb Z$ acting as $\tau\sim\tau+\beta$ on the cylinder of \eqref{eq:cylinder-metric}.  Let $\gamma_\beta$ denote this non-contractible thermal circle. We will keep the global cylinder Weyl frame $R_\partial$ arbitrary\footnote{Keeping $R_\partial$ explicit distinguishes the arbitrary conformal scale of the chosen boundary-cylinder representative from the Cartan normalization scale $\ell$ (which eventually becomes the AdS radius).}. The thermal conformal data depend on the dimensionless modulus $\beta/R_\partial$.

The conformal representation labels $(\Delta,J,\ldots)$ may be specified directly in the Euclidean theory.  When the same conformal family is interpreted as a physical Hilbert-space representation after Lorentzian continuation, it becomes a positive-energy lowest-weight representation of $SO(d,2)$. The Euclidean thermal loop below computes the corresponding conformal character, or equivalently, the ordinary thermal trace on the CFT cylinder.

Along $\gamma_\beta$, the cylinder connection~\eqref{eq:a-cylinder} reduces to
\beq
 a_\tau
 =\frac1\ell P_0+\frac{\ell}{4R_\partial^2}K_0
 \defi\frac1{R_\partial}H_\partial\,.
\label{eq:thermal-Hpartial}
\eeq
In radial gauge the holonomy is therefore
\beq
U_\beta(\rho)
\defi {\cal P}\exp\!\oint_{\gamma_\beta}A
=b(\rho)^{-1}
\exp\!\left(\frac{\beta}{R_\partial}H_\partial\right)b(\rho)
\label{eq:thermal-holonomy-radial}
\eeq
and is independent of $\rho$ up to conjugation, as expected for a flat connection.

The conjugacy class of $H_\partial$ is particularly simple.  Set
\beq
s_\partial\defi\log\frac{2R_\partial}{\ell}\,,\qquad
X\defi\frac12(P_0+K_0)\,.
\label{eq:thermal-conjugacy-defs}
\eeq
Using~\eqref{eq:conf-algebra-1},
\beq
e^{s_\partial D}H_\partial e^{-s_\partial D}=X\,.
\label{eq:Hpartial-to-X}
\eeq
Meanwhile~\eqref{eq:conf-algebra-1} -- \eqref{eq:conf-algebra-3} imply
\[
[P_0-K_0,D]=2X\,,\qquad [P_0-K_0,X]=-2D\,,
\]
so that
\beq
\exp\!\left[\frac{\pi}{4}(P_0-K_0)\right]
D
\exp\!\left[-\frac{\pi}{4}(P_0-K_0)\right]
=X\,.
\label{eq:X-conjugate-D}
\eeq
Thus $H_\partial$ is conjugate to $D$, and consequently
\beq
U_\beta\ \sim\ \exp\!\left(\frac{\beta}{R_\partial}D\right)
\label{eq:thermal-holonomy-D}
\eeq
where $\sim$ denotes equality of conjugacy classes.  This is the Cartan-connection version of the state-operator map: under the conformal map from the cylinder to the plane, $\tau\to\tau+\beta$ becomes a dilation by $e^{\beta/R_\partial}$.

For a closed loop it is convenient to distinguish the traced Wilson observable from the parallel transport $W_R$ of~\eqref{eq:BF-electric-line-bootstrap}:
\beq
{\cal W}_R(\gamma)
\defi \Tr_R W_R(\gamma)
=\Tr_R\,{\cal P}\exp\!\left(\oint_\gamma A\right).
\label{eq:traced-wilson-def}
\eeq
Let $V_{\Delta,0}$ be a generic long\footnote{The character and Fock-space arguments used below extend directly to shortened representations, provided $V$ is taken to be the physical (irreducible) module, so that its character includes the appropriate subtraction of null descendants.  What becomes more involved is the explicit decomposition of $\operatorname{Sym}^N V$ into irreducible conformal modules.  Carrying out this decomposition for the stress-tensor module would give the corresponding multi-stress-tensor Wilson tower, and is of direct interest from the bulk gravity perspective.} scalar lowest-weight representation of $SO(2,d)$. Because our convention is $[D,P_i]=-P_i$, the positive cylinder Hamiltonian is $\widehat D=-D$: a scalar primary obeys
\beq
D|\Delta\rangle=-\Delta|\Delta\rangle\,,\qquad
K_i|\Delta\rangle=0\,.
\label{eq:D-primary-convention}
\eeq
Its level-$n$ descendants
\[
P_{i_1}\cdots P_{i_n}|\Delta\rangle
\]
have $D$-eigenvalue $-(\Delta+n)$ and multiplicity
$\binom{n+d-1}{d-1}$.  Defining
\beq
q\defi e^{-\beta/R_\partial}\,,
\label{eq:q-thermal-def}
\eeq
we obtain
\beq
\begin{aligned}
{\cal W}_{\Delta,0}(\gamma_\beta)
&=\Tr_{V_{\Delta,0}}\exp\!\left(\frac{\beta}{R_\partial}D\right)
\\
&=\sum_{n=0}^\infty
\binom{n+d-1}{d-1}q^{\Delta+n}
=\frac{q^\Delta}{(1-q)^d}
\equiv \chi_{\Delta,0}(q)\,.
\end{aligned}
\label{eq:scalar-wilson-character}
\eeq
Thus the traced thermal Wilson loop is the global conformal character of the scalar module.

There is an independent bulk field calculation, that has connections to this. In the AdS specialization, the same $\ell$ is the physical AdS radius. A scalar with
\beq
m^2\ell^2=\Delta(\Delta-d)
\label{eq:scalar-mass-dimension}
\eeq
has global-AdS normal-mode energies $\ell E_{n,s}=\Delta+2n+s$.  The degeneracy $g_s^{(d)}$ of scalar spherical harmonics of angular momentum $s$ on $S^{d-1}$ obeys
\beq
\sum_{s=0}^\infty g_s^{(d)}q^s
=\frac{1-q^2}{(1-q)^d}\,.
\label{eq:spherical-harmonic-generating}
\eeq
Setting $R_\partial=\ell$ (the global-AdS frame, in which the boundary cylinder radius coincides with the physical AdS radius),
\beq
Z_{\rm 1p}(q)
=\sum_{n,s\geq0}g_s^{(d)}q^{\Delta+2n+s}
=\frac{q^\Delta}{(1-q)^d}
={\cal W}_{\Delta,0}(\gamma_\beta)\,.
\label{eq:one-particle-equals-wilson}
\eeq
The Wilson loop and the one-particle partition function are therefore the same $SO(2,d)$ trace.

\subsection{The One-Loop Determinant as a Multi-Winding Wilson Spool} \label{subsec:higher-d-wilson-spool}

The thermal part of the free scalar one-loop determinant on thermal AdS$_{d+1}$ is known from
heat-kernel, image-sum and normal-mode methods
\cite{GopakumarGuptaLalHeatKernel,KakkarSarkarAdS}.  With the vacuum contribution divided out, it is
\beq
\log Z^{(1)}_{\Delta,\rm th}(q)
=\sum_{k=1}^\infty\frac1k
\frac{q^{k\Delta}}{(1-q^k)^d}\,.
\label{eq:thermal-scalar-one-loop}
\eeq
Using~\eqref{eq:scalar-wilson-character}, this becomes
\beq
\log Z^{(1)}_{\Delta,\rm th}
=\sum_{k=1}^\infty\frac1k\,
{\cal W}_{\Delta,0}(\gamma_\beta^k)
=\sum_{k=1}^\infty\frac1k\,
\Tr_{V_{\Delta,0}}U_\beta^k\, .
\label{eq:higher-d-wilson-spool}
\eeq
The integer $k$ is literally the winding number around the non-contractible $S^1_\beta$. Equation~\eqref{eq:higher-d-wilson-spool} is therefore the higher-dimensional analogue of a Wilson spool.

The relation also makes transparent why a one-loop determinant appears.  The bosonic Fock space generated by the one-particle module is
\beq
{\cal F}_{\rm B}(V_{\Delta,0})
=\bigoplus_{N=0}^\infty \operatorname{Sym}^N V_{\Delta,0}\,.
\label{eq:bosonic-fock-symmetric}
\eeq
The standard plethystic identity gives
\beq
Z^{(1)}_{\Delta,\rm th}
=\exp\!\left[
\sum_{k=1}^\infty\frac1k\,
\Tr_{V_{\Delta,0}}U_\beta^k
\right]
=\sum_{N=0}^\infty
\Tr_{\operatorname{Sym}^N V_{\Delta,0}}U_\beta\,.
\label{eq:plethystic-fock-wilson}
\eeq
Thus the winding sum is simply the cycle decomposition of the second-quantized trace.

\subsection{The Multi-Trace Wilson Tower}
\label{subsec:multitrace-wilson-tower}

The last expression in~\eqref{eq:plethystic-fock-wilson} is the sum over characters of the $N$-particle bosonic Hilbert space. We want to decompose the $N$-particle space into irreducible global conformal modules. In radial quantization each such module is generated from a conformal primary.  In the generalized-free limit, the primaries appearing in $\operatorname{Sym}^N V_{\Delta,0}$ are the $N$-trace primaries built from the single-trace operator ${\cal O}$ dual to the bulk field \cite{HeemskerkPenedonesPolchinskiSully,FitzpatrickKaplanAdSField}.  Concretely, an $N$-trace primary is a linear combination of local composites made from $N$ copies of ${\cal O}$ and derivatives, with descendants removed so that the resulting operator is annihilated by the special conformal generators.\footnote{Here $N$ counts particles, also called the number of ``traces", in the generalized-free CFT sector. We will not explicitly connect it to any gauge theory trace. Also, this $N$ should not (of course) be thought of as the $N$ in large-$N$. That $N$ is infinite in this section.}  We write
\beq
\operatorname{Sym}^N V_{\Delta,0}
=\bigoplus_{{\cal P}\in{\rm MT}_N({\cal O})}
 m_{\cal P}\,V_{\Delta_{\cal P},J_{\cal P}}\,.
\label{eq:symmetric-power-primary-decomposition}
\eeq
Here ${\rm MT}_N({\cal O})$ denotes the set of $N$-trace conformal primaries and
$m_{\cal P}$ their multiplicities.

The symmetric-power character on the left of
\eqref{eq:symmetric-power-primary-decomposition} is obtained directly from the one-particle
character by the standard cycle-index formula \cite{MacdonaldSymmetricFunctions,BenvenutiPlethystics}.  If
$\lambda=(1^{m_1}2^{m_2}\cdots)$ is a partition of $N$ and
$z_\lambda=\prod_k k^{m_k}m_k!$, then
\beq
\Tr_{\operatorname{Sym}^N V}U
=\sum_{\lambda\vdash N}\frac1{z_\lambda}
\prod_{k\geq1}\big(\Tr_V U^k\big)^{m_k}\,.
\label{eq:cycle-index-symmetric-power}
\eeq
For the first two nontrivial cases,
\beq
\begin{aligned}
\Tr_{\operatorname{Sym}^2 V}U
&=\frac12\big[(\Tr_VU)^2+\Tr_VU^2\big]\,,\\
\Tr_{\operatorname{Sym}^3 V}U
&=\frac16\big[(\Tr_VU)^3+3(\Tr_VU)(\Tr_VU^2)+2\Tr_VU^3\big] .
\end{aligned}
\label{eq:sym2-sym3-cycle-index}
\eeq
The terms on the right are the cycle types of the permutation projector onto the symmetric power.  In the Wilson-loop language they are correspondingly products of once-wound loops and multiply wound loops. The primary content is obtained only in the second step, by re-expanding this character in irreducible conformal characters.  This is the standard plethystic/character logic used in operator counting \cite{DolanConformalCharacters,HenningOperatorBases}. In the present setting each character also has the conformal-Cartan Wilson-loop realization of \eqref{eq:traced-wilson-def} -- \eqref{eq:scalar-wilson-character}.  The analogous symmetric-function reorganization in AdS$_3$ was used in~\cite{Vishal}.

The descendant factor is useful in making the second step explicit.  Introduce maximal-torus fugacities\footnote{The variables $\mathbf y$ are auxiliary angular fugacities that refine the thermal character by keeping track of the $SO(d)$ representation carried by each state: writing $y_a=e^{i\theta_a}$, one replaces $\Tr_V q^{\widehat D}$ by $\Tr_V\!\left(q^{\widehat D}\prod_a y_a^{J_a}\right)$, with $J_a$ Cartan generators of $SO(d)$.  Thus $\chi_{\Delta,J}(q,\mathbf y)=\Tr_{\mathcal V_{\Delta,J}}\!\left(q^{\widehat D}\prod_a y_a^{J_a}\right)$ is the full conformal character of the multiplet $\mathcal V_{\Delta,J}$, while $\chi^{SO(d)}_J(\mathbf y)=\Tr_J\!\left(\prod_a y_a^{J_a}\right)$ is the finite-dimensional $SO(d)$ character of the primary representation $J$.  They are useful here because the refined character retains the spin information needed to decompose a symmetric power into conformal multiplets. Setting $\mathbf y=\mathbf 1$ recovers the unrefined thermal trace used above.} $\mathbf y=(y_1,\ldots,y_r)$ for $SO(d)$, with $r=\lfloor d/2\rfloor$.  For a generic long conformal module, the translations generate the universal descendant factor \cite{DolanConformalCharacters,HenningOperatorBases}
\beq
{\cal P}_d(q,\mathbf y)=
\begin{cases}
\displaystyle\prod_{a=1}^{r}
\frac1{(1-qy_a)(1-qy_a^{-1})}\,,&d=2r\,,\\[4mm]
\displaystyle\frac1{1-q}\prod_{a=1}^{r}
\frac1{(1-qy_a)(1-qy_a^{-1})}\,,&d=2r+1\,.
\end{cases}
\label{eq:conformal-descendant-factor-refined}
\eeq
Accordingly, for a long primary of dimension $\Delta$ and $SO(d)$ representation $J$,
\beq
{\cal W}_{\Delta,J}(q,\mathbf y)
=\chi_{\Delta,J}(q,\mathbf y)
=q^\Delta\chi^{SO(d)}_J(\mathbf y)\,{\cal P}_d(q,\mathbf y)\,.
\label{eq:refined-long-wilson-character}
\eeq
For the scalar one-particle module, $\chi_{\Delta,0}(q,\mathbf y)=q^\Delta{\cal P}_d(q,\mathbf y)$, and in the cycle-index formula $\Tr_VU^k=\chi_{\Delta,0}(q^k,\mathbf y^k)$.  Therefore, once the right-hand side of \eqref{eq:cycle-index-symmetric-power} has been assembled, the multi-trace primaries can be read off by expanding in the basis $q^{\Delta_{\cal P}}\chi^{SO(d)}_{J_{\cal P}}(\mathbf y){\cal P}_d(q,\mathbf y)$.  Equivalently, for generic $\Delta$ for which the multi-trace modules are long, one may divide out the common descendant factor:
\beq
{\cal G}^{\rm prim}_N(q,\mathbf y)
\defi
\frac{\Tr_{\operatorname{Sym}^N V_{\Delta,0}}U}{
{\cal P}_d(q,\mathbf y)}
=
\sum_{{\cal P}\in{\rm MT}_N({\cal O})}
 m_{\cal P}\,q^{\Delta_{\cal P}}
 \chi^{SO(d)}_{J_{\cal P}}(\mathbf y)\,.
\label{eq:Ntrace-primary-generating}
\eeq
This quotient is the generating function for the (long) primary spectrum itself: the powers of $q$ give the primary dimensions, the $SO(d)$ characters give their spins, and the
coefficients give their multiplicities.  

Taking the thermal trace of~\eqref{eq:symmetric-power-primary-decomposition} and using
\eqref{eq:traced-wilson-def} gives
\beq
\Tr_{\operatorname{Sym}^N V_{\Delta,0}}U_\beta
=\sum_{{\cal P}\in{\rm MT}_N({\cal O})}
 m_{\cal P}\,{\cal W}_{\Delta_{\cal P},J_{\cal P}}(\gamma_\beta)\,.
\label{eq:Ntrace-as-single-wilson}
\eeq
Every Wilson loop on the right winds the thermal circle only once.  Combining this with
\eqref{eq:plethystic-fock-wilson} gives
\beq
Z^{(1)}_{\Delta,\rm th}
=1+\sum_{N=1}^\infty
\sum_{{\cal P}\in{\rm MT}_N({\cal O})}
 m_{\cal P}\,
 {\cal W}_{\Delta_{\cal P},J_{\cal P}}(\gamma_\beta)\, .
\label{eq:one-loop-wilson-tower}
\eeq
Thus the multi-winding Wilson spool and the single-winding multi-trace Wilson tower are two character decompositions of the same free Fock-space trace.  The cycle-index expressions such as~\eqref{eq:sym2-sym3-cycle-index} provide one key step: the conformal-character re-expansion then extracts the actual primary modules. 

Some aspects of this decomposition are more direct in $d>2$ than in the $d=2$ analysis of \cite{Vishal}.  In two dimensions one must keep track of holomorphic and anti-holomorphic descendants separately, while imposing bosonic permutation symmetry on the full left-right state. This leads to additional complications. Here the $SO(d)$ fugacities package the spin content directly into ordinary rotation-group characters.

\subsection{Illustration: The Double-Trace Sector}
\label{subsec:double-trace-wilson-check}

The $N=2$ case makes the above logic explicit.  Substituting the scalar character $\chi_{\Delta,0}=q^\Delta{\cal P}_d$ into the first line of \eqref{eq:sym2-sym3-cycle-index} gives the full two-particle character
\beq
\chi_{\operatorname{Sym}^2V_{\Delta,0}}(q,\mathbf y)
=\frac{q^{2\Delta}}2\left[
{\cal P}_d(q,\mathbf y)^2
+{\cal P}_d(q^2,\mathbf y^2)
\right].
\label{eq:double-trace-cycle-refined}
\eeq
Expanding this expression directly in powers of $q$ would only count two-particle states. To isolate conformal primaries, divide by the universal descendant factor as in \eqref{eq:Ntrace-primary-generating}. It can be checked (for both even and odd $d$) that
\beq
\frac{{\cal P}_d(q^2,\mathbf y^2)}{{\cal P}_d(q,\mathbf y)}
={\cal P}_d(-q,\mathbf y)\,,
\label{eq:Pd-cycle-ratio}
\eeq
from which we obtain
\beq
{\cal G}^{\rm prim}_2(q,\mathbf y)
=\frac{q^{2\Delta}}2\left[
{\cal P}_d(q,\mathbf y)+{\cal P}_d(-q,\mathbf y)
\right].
\label{eq:double-trace-primary-generating}
\eeq
Now use the standard generating function for rank-$s$ symmetric-traceless $SO(d)$
representations,
\beq
(1-q^2){\cal P}_d(q,\mathbf y)
=\sum_{s=0}^\infty q^s\chi^{SO(d)}_s(\mathbf y)\,.
\label{eq:stt-generating-function}
\eeq
The average of the $q$ and $-q$ expressions projects onto even spin, so
\beq
\frac12\left[{\cal P}_d(q,\mathbf y)+{\cal P}_d(-q,\mathbf y)\right]
=\frac1{1-q^2}\sum_{s=0,2,4,\ldots}
 q^s\chi^{SO(d)}_s(\mathbf y)
=\sum_{n=0}^\infty\sum_{s=0,2,4,\ldots}
 q^{2n+s}\chi^{SO(d)}_s(\mathbf y)\,.
\label{eq:even-spin-primary-generating}
\eeq
Combining~\eqref{eq:double-trace-primary-generating} and \eqref{eq:even-spin-primary-generating} therefore reads off the primary content directly: there is one primary for every $n\geq0$ and even $s\geq0$, with
\beq
\Delta_{n,s}=2\Delta+2n+s\,.
\label{eq:double-trace-dimensions}
\eeq
These are the familiar generalized-free double-trace primaries $[\mathcal O\mathcal O]_{n,s}$ \cite{HeemskerkPenedonesPolchinskiSully,FitzpatrickKaplanAdSField}.  Schematically they are bilinears $\mathcal O\,\partial_{\{\mu_1}\!\cdots\partial_{\mu_s\}}(\partial^2)^n\mathcal O+\cdots$, where the omitted terms remove traces and conformal descendants.  Thus the remarkable cycle-index formula has not merely counted the two-particle states: after stripping the common descendant factor, it has produced their decomposition into conformal-primary modules.

Multiplying the primary generating function back by ${\cal P}_d$ gives the module decomposition
\beq
\operatorname{Sym}^2V_{\Delta,0}
=\bigoplus_{n=0}^\infty
\bigoplus_{s=0,2,4,\ldots}
V_{2\Delta+2n+s,\,s}\,.
\label{eq:double-trace-module-decomposition}
\eeq
Replacing the characters by their thermal Wilson traces gives the corresponding Wilson-loop identity,
\beq
\frac12\left[
{\cal W}_{\Delta,0}(\gamma_\beta)^2
+{\cal W}_{\Delta,0}(\gamma_\beta^2)
\right]
=\sum_{n=0}^\infty\sum_{s=0,2,4,\ldots}
{\cal W}_{2\Delta+2n+s,\,s}(\gamma_\beta)\, .
\label{eq:double-trace-wilson-identity}
\eeq
At zero angular fugacity, $\chi^{SO(d)}_s(\mathbf 1)=d_s^{(d)}$ and ${\cal P}_d(q,\mathbf 1)=(1-q)^{-d}$, so the same statement becomes
\beq
\sum_{n\geq0}\sum_{s\ {\rm even}}
\frac{d_s^{(d)}q^{2\Delta+2n+s}}{(1-q)^d}
=\frac12\left[
\frac{q^{2\Delta}}{(1-q)^{2d}}
+\frac{q^{2\Delta}}{(1-q^2)^d}
\right].
\label{eq:double-trace-character-check}
\eeq
This last equality is a useful unrefined check, but the refined derivation above is the important step: it shows explicitly how the right-hand side of the $N=2$ cycle-index formula is reorganized into the dimensions, spins and multiplicities of the double-trace primaries.

\subsection{Bulk Light Fields and the Meaning of the Tower}
\label{subsec:general-light-field-tower}

Nothing essential in the preceding argument is special to scalars.  Let $V$ be the
one-particle $SO(2,d)$ module of a perturbative bulk field.  For a boson or fermion one has,
respectively,
\beq
\log Z^{(1)}_{V,\rm th}
=\sum_{k=1}^\infty\frac{\eta^{k+1}}{k}\,
\Tr_V U_\beta^k
=\sum_{k=1}^\infty\frac{\eta^{k+1}}{k}\,
{\cal W}_V(\gamma_\beta^k)\,,
\qquad
\eta=\begin{cases}+1,&\text{boson},\\-1,&\text{fermion}.\end{cases}
\label{eq:boson-fermion-wilson-spool}
\eeq
The Fock sectors are $\operatorname{Sym}^N V$ for bosons and $\wedge^N V$ for fermions.  Decomposing these into irreducible conformal modules again produces a tower of single-winding Wilson loops. For gauge fields or other shortened representations, $V$ should be understood as the physical irreducible one-particle module, whose character contains the appropriate null-state subtractions.  Such shortened characters appear explicitly in character formulations of higher-spin AdS one-loop determinants \cite{GopakumarGuptaLalHeatKernel,SunAdSCharacters}.  The same plethystic construction then applies to this physical character.  The additional nontrivial step, which we will not attempt here, is to re-expand the resulting multi-particle characters explicitly into irreducible conformal modules while keeping track of the shortening relations.

The point of these observations is to emphasize that the existence of an underlying bulk TQFT skeleton need not imply that the bulk itself is topological\footnote{There are two uses of the word ``bulk" in this sentence. The first refers to the (microscopic) topological bulk, and the second, to the bulk in EFT which we call spacetime.}: we can have multi-trace towers of bootstrapped Wilson line sectors, which can look like local bulk degrees of freedom. In fact, we believe the structure is fundamentally the same in $d=2$ and $d > 2$, even though AdS$_3$ gravity is often characterized as a ``toy" model. Our calculation here is conceptually the same as the one done in \cite{Vishal}, even though there are some technical differences. The main distinction in higher dimensions is the presence of a stress tensor that is not a (Virasoro) descendant of the vacuum. The Virasoro vacuum character in $d=2$ in effect contains the entire tower of multi-stress tensor primaries and their (global) descendants. But this does not invalidate the fact that other light fields can nonetheless propagate in AdS$_3$. The distinction is simply that pure gravity in 3d is closer to the topological skeleton than it is in higher dimensions \cite{Vishal2}.

Of course, the sector we have considered here is best viewed as a decoupled free sector of the full CFT. In the full theory, light bulk fields will have  interactions suppressed by large-$N$. But we do not believe that this changes the substantive picture that bulk local fields are best viewed as (interacting) multi-trace towers of Wilson lines in an ultimately gluing bootstrap consistent spectrum.

This also differs in philosophy from spin-foam approaches where local degrees of freedom are introduced by demanding ``simplicity" as an explicit constraint on the $B$-field, directly at the level of the BF theory. We will have more to say about it in the next section, where we will see how simplicity emerges as a type of effective constraint on bootstrapped CFTs with an Einstein gravity dual.

\section{Generalized Simplicity and Semi-Classical AdS/CFT}
\label{sec:ward-simplicity}

Section~\ref{subsec:wilson-line-center} described an electric Wilson representation using its linking $B$-flux.  On the boundary, the spacetime SymTFT gives this flux a physical interpretation: the relevant $B$-currents are the conformal currents built from the stress tensor.  We will call this local relation \emph{generalized simplicity}.  It provides a physical interpretation of the conformal BF data that makes sense independently of whether the CFT has a semiclassical gravitational dual.

When the CFT admits a semiclassical AdS regime described by a local gravitational EFT, the same stress-tensor sector has a realization in terms of bulk gravitational variables, through the semi-classical AdS/CFT correspondence. This suggests a natural definition of an {\em effective} $B$-field via the effective stress tensor. For Einstein gravity this reduces to the so-called ``simplicity" relation (used in spon-foam models), while more general gravitational EFTs lead to generalizations. The purpose of this section is to develop this connection.

\subsection{Conformal $B$-flux and the Stress Tensor}
\label{subsec:generalized-simplicity-current}

The relation between the conformal $B$-field and the CFT stress tensor is easiest to see first in the flat-space background used explicitly by Apruzzi, Dondi, Garc\'ia-Etxebarria, Lam and Sch\"afer-Nameki ~\cite{ApruzziSpacetimeSymTFT}. They consider a Dirichlet boundary condition that fixes the boundary value of a flat BF connection $A$. In Section~3.2 of that work they make a more specific choice, which we write as
\beq
 \bar A_{\rm flat}=dx^a P_a\,,
\label{eq:Apruzzi-flat-Dirichlet}
\eeq
which is the BF-flat conformal Cartan connection of flat Euclidean space in their conventions.  In the notation of our Section~\ref{subsec:conformal-radial}, this is simply the flat representative (in the conformally flat class) of the BF-flat conformal Cartan connection,
\beq
 a[\delta]=\frac1\ell dx^aP_a\,,
\eeq
up to our explicit normalization of $P_a$ by $\ell$. Thus the special choice~\eqref{eq:Apruzzi-flat-Dirichlet} should be viewed as the flat-metric representative within the broader class of BF-flat conformal Cartan connections (equivalently, locally conformally flat boundary conformal structures) allowed by the BF equations.

Following Apruzzi et al., we use the Killing form to identify $\mathfrak g\simeq\mathfrak g^*$ to write $B$ in the dual conformal basis as
\beq
B
=
\frac12\,\mathfrak j^{ab}M^*_{ab}
+t^aP^*_a+s^aK^*_a+\varphi D^*\,.
\label{eq:B-conformal-dual-decomposition}
\eeq
Locally, they fix the $B$-shift gauge symmetry using
\beq
\mathfrak j^{ab}=s^a=\varphi=0\,.
\label{eq:Apruzzi-B-shift-gauge}
\eeq
The non-trivial components of $D_{\bar A_{\rm flat}}B=0$ then reduce to
\beq
 e^{[a}\wedge t^{b]}=0\,,
\qquad
dt^a=0\,,
\qquad
e_a\wedge t^a=0\,,
\label{eq:B-equations-stress-Ward}
\eeq
with $e^a=dx^a$.  Identifying\footnote{Here and below, $\star_{\partial}$ denotes the Hodge dual of the boundary metric.}
\beq
-\frac{1}{\kappa_{\rm BF}}\,t^a=\star_{\partial}\tau^a\,,
\qquad
\tau^a\defi e^a{}_{\mu}T^{\mu\nu}dx_\nu\,,
\label{eq:t-a-stress-tensor}
\eeq
these equations say, respectively,
\beq
T_{\mu\nu}=T_{\nu\mu}\,,
\qquad
\partial_\mu T^{\mu\nu}=0\,,
\qquad
T^\mu{}_{\mu}=0\,.
\label{eq:flat-stress-Ward}
\eeq
This is the main physical point of the calculation: after the local $B$-shift redundancy is removed, the BF equation packages the symmetry, conservation and tracelessness Ward identities of a CFT stress tensor.  

The translation, rotation, dilatation and special-conformal charges constructed from the charge-normalized $B$-field, $-B/\kappa_{\rm BF}$, then become the standard zeroth, first and second $x$-moments\footnote{When referring to the explicit ``$x$-moments'' of the stress tensor, $x^a$ denotes the flat boundary coordinates adapted to the representative $\bar A_{\rm flat}=dx^aP_a$, as in \cite{ApruzziSpacetimeSymTFT}.  By contrast, in expressions such as $\zeta_X(x)$, $x$ denotes a generic boundary point (or local coordinates there): no choice of flat representative is implied. We don't expect that this will cause any confusion.} of $T_{\mu\nu}$ \cite{ApruzziSpacetimeSymTFT}.  Rather than tying the construction to the flat representative $\bar A_{\rm flat}=dx^a P_a$, we would like a formulation that continues to apply in any conformally flat representative.  The natural way to do this is to start from a fixed conformal generator $X\in\mathfrak g$ at some reference point on the boundary, and parallel transport it using the flat connection $A$, to define it at other points.  Locally this can always be done. Whenever the final current is integrated over an extended or closed region below, it is understood that the corresponding parallely transported section is defined there.  This gives an adjoint-valued field $\zeta_X(x)$ representing the same generator in the local frame at each point. By construction it is covariantly constant,
\beq
D_A\zeta_X
\equiv d\zeta_X+[A,\zeta_X]=0.
\label{eq:zetaX-parallel}
\eeq
On a simply connected patch, where $A=u^{-1}du$, fixing the frame at the reference point gives $\zeta_X=u^{-1}Xu$.  Although $X$ itself is fixed, its decomposition into translation, rotation, dilatation and special-conformal generators therefore becomes position dependent.  In the flat representative~\eqref{eq:Apruzzi-flat-Dirichlet}, this position dependence reproduces precisely the constant, linear and quadratic combinations of $x$ that enter the corresponding stress-tensor moment currents of \cite{ApruzziSpacetimeSymTFT}.  

To obtain the covariant statement, we now use the fact that BF-flatness implies conformal flatness. So the connection we are interested in, is locally gauge-equivalent to~\eqref{eq:Apruzzi-flat-Dirichlet}. The BF equations and the $B$-shift symmetry are gauge covariant, so the complete flat-space construction above can be transported to the curved conformally-flat Cartan frame. At the same time, (in the anomaly-free cases) the CFT Ward identities also transform covariantly. Thus the transported tensor obeys
\beq
T_{\mu\nu}=T_{\nu\mu}\,,
\qquad
\nabla_\mu T^{\mu\nu}=0\,,
\qquad
T^\mu{}_{\mu}=0\,.
\label{eq:curved-stress-Ward}
\eeq
For a generator $X$, let $\xi_X$ denote the corresponding conformal Killing vector and define the associated CFT current
\beq
j_X[T]\defi T_{\mu\nu}\xi_X^\nu dx^\mu\,,
\qquad
{\cal J}_X[T]\defi\star_{\partial}j_X[T]\in\Omega^{d-1}(\partial M)\,.
\label{eq:stress-current-form}
\eeq
Equation~\eqref{eq:curved-stress-Ward} and the conformal Killing equation imply
\beq
\nabla_\mu\!\left(T^\mu{}_{\nu}\xi_X^\nu\right)
=
T^{\mu\nu}\nabla_{(\mu}\xi_{X\nu)}
=
\frac1d(\nabla\!\cdot\!\xi_X)T^\mu{}_{\mu}
=0\,,
\eeq
so $d{\cal J}_X[T]=0$ away from insertions.  The BF field $B$ carries an analogous conserved-current datum.  Pairing it with the covariantly constant adjoint field $\zeta_X$ gives the ordinary $(d-1)$-form $\langle\zeta_X,B\rangle$, and the equations $D_A\zeta_X=0$ and $D_AB=0$ imply that this form is closed,\footnote{The invariant bilinear form obeys $\langle [A,\zeta_X],B\rangle+\langle \zeta_X,[A,B]\rangle=0$. Hence the connection terms cancel in the contraction, so that $d\langle \zeta_X,B\rangle =\langle D_A\zeta_X,B\rangle+\langle \zeta_X,D_AB\rangle$.}
\beq
d\langle\zeta_X,B\rangle=0\,.
\eeq
Under the shift symmetry $B\mapsto B+D_A\Xi$,
\beq
\langle\zeta_X,B\rangle
\ \longmapsto\
\langle\zeta_X,B\rangle+d\langle\zeta_X,\Xi\rangle\,.
\label{eq:B-current-exact-shift}
\eeq
Thus the gauge-invariant content of $\langle\zeta_X,B\rangle$ is its conserved-current class modulo exact terms, precisely paralleling the improvement ambiguity of ${\cal J}_X[T]$. The natural identification is therefore
\beq
\big[-\frac{1}{\kappa_{\rm BF}}\langle\zeta_X,B\rangle\big]
\ \longleftrightarrow\
\big[{\cal J}_X[T]\big]\,,
\label{eq:BF-CFT-current-class}
\eeq
where the brackets denote conserved currents modulo exact improvements. 

The covariant form of the identification can now be written as
\beq
\left[
-\frac{1}{\kappa_{\rm BF}}\langle\zeta_X,B\rangle\Big|_{\partial M}
\right]_{\rm imp}
=
\left[{\cal J}_X[T]\right]_{\rm imp}\,,
\qquad X\in\mathfrak{so}(d+1,1)\,.
\label{eq:generalized-simplicity-cohomology}
\eeq
Here $[\,\cdot\,]_{\rm imp}$ denotes the on-shell current class modulo local exact improvements. Equivalently, local representatives can be chosen so that
\beq
-\frac{1}{\kappa_{\rm BF}}\langle\zeta_X,B\rangle\Big|_{\partial M}
=
\star_{\partial}\!\left(T_{\mu\nu}\xi_X^\nu dx^\mu\right)+d\chi_X\,.
\label{eq:generalized-simplicity-local}
\eeq
This is the covariant form of the Apruzzi et al. relation that we will work with.

Integrating over a closed cut $\Sigma$ removes the improvement term and gives the corresponding charge equality,
\beq
\mathcal Q_X^{\rm BF}[\Sigma]
\defi
-\frac{1}{\kappa_{\rm BF}}
\int_\Sigma\langle\zeta_X,B\rangle
=
\int_\Sigma {\cal J}_X[T]
=
\int_\Sigma d\Sigma_\mu\,T^\mu{}_{\nu}\xi_X^\nu
\defi
\mathcal Q_X^{\rm CFT}[\Sigma]\,.
\label{eq:normalized-BF-charge}
\eeq
This is also the form that directly connects to the Wilson-line discussion of Section~\ref{subsec:wilson-line-center}: the linking $B$-flux that labels the Wilson representation is the corresponding CFT conformal charge. This is what makes the identification of the corresponding states as bulk eigenstates.

Finally, the sourced BF equation gives the local Ward identity in the presence of a charged insertion.  Pairing~\eqref{eq:sourced-BF} with $\zeta_X$ gives
\beq
d\!\left[-\frac{1}{\kappa_{\rm BF}}\langle\zeta_X,B\rangle\right]
=
\langle\zeta_X,J_C\rangle\,.
\label{eq:BF-local-Ward-source}
\eeq
When the Wilson line ends on, or is identified with, a CFT operator in representation ${\cal R}$, the right-hand side inserts the action of $X$ on ${\cal R}$.  Integrating over a small ball around the endpoint therefore reproduces the usual conformal Ward identity: the flux through its linking sphere measures the conformal charge of the operator.

We will call the local current identification~\eqref{eq:generalized-simplicity-cohomology} \emph{generalized simplicity}.  Its non-trivial content is that the physical CFT boundary realizes the different conformal $B$-currents through one and the same symmetric, conserved and traceless rank-two primary operator,
\beq
{\mathfrak S}_{\rm CFT}:
T_{\mu\nu}
\ \longmapsto\
\left\{
\left[\star_{\partial}(T_{\mu\nu}\xi_X^\nu dx^\mu)\right]_{\rm imp}
\right\}_{X\in\mathfrak{so}(d+1,1)}\,.
\label{eq:CFT-generalized-simplicity-map}
\eeq
The CFT stress-tensor currents are fixed by the Ward identities. Their identification with the BF currents is the physical boundary-current identification above and does not assume (yet) an Einstein gravitational dual. In even boundary dimension, conformal anomalies can modify the Weyl-transformed Ward identities and require the additional Chern-Simons terms of the full spacetime SymTFT \cite{ApruzziSpacetimeSymTFT}. This is as an interesting, but (for our purposes) secondary, complication: we leave it for later.

\subsection{EFT, Einstein Gravity, Simplicity}
\label{subsec:Einstein-emergent-simplicity}

Now we restrict to a CFT for which the established AdS/CFT dictionary admits a weakly coupled, local gravitational EFT.  Familiar diagnostics of this regime include large $C_T$ (controlled by $N$), large-$N$ factorization, a sparse set of low-dimension single-trace operators, a parametrically large gap to single-trace higher-spin states, and sufficiently soft Regge behavior.  Large $N$ together with a large gap organizes crossing solutions into local AdS interactions \cite{HeemskerkPenedonesPolchinskiSully,FitzpatrickKaplanAdSField}, while Regge causality and the higher-spin gap suppress non-Einstein graviton couplings \cite{CamanhoEdelsteinMaldacenaZhiboedov,AfkhamiJeddiHartmanKunduTajdini,MeltzerPerlmutter}.  We assume a single interacting stress-tensor sector in the bootstrapped CFT, in what follows.\footnote{The purpose here is to identify the BF/gravity dictionary in a regime where a local gravitational EFT is already supplied by the usual AdS/CFT correspondence. A genuine derivation of effective bulk dynamics within the present paradigm would instead require directly working with the bootstrapped Wilson lines (especially the multi-stress tensor tower of Wilson lines) interacting via their trivalent couplings. We expect that this will lead to effective metric dynamics we are familiar from the bulk. The key point is that this is in principle tractable: it gives us bulk ingredients to work with, instead of (at some stage) postulating the bulk spacetime, as is usually done in most approaches to bulk reconstruction.}

In this regime $T_{\mu\nu}$ is the operator dual to the massless spin-two field.  In vielbein
variables the renormalized generating functional obeys
\beq
\delta W_{\rm CFT}[e_{(0)}]
=
\int_{\partial M}d^dx\,|e_{(0)}|\,
\langle T^\mu{}_a\rangle\,\delta e_{(0)\mu}{}^a\,.
\label{eq:holographic-vielbein-source}
\eeq
Thus the boundary coframe is the source for $T_{\mu\nu}$ \cite{BalasubramanianKraus,deHaroSkenderisSolodukhin}. As usual in holographic renormalization, the stress tensor is identified with the finite renormalized canonical momentum after the standard boundary terms and counterterms -- not directly with the bare divergent asymptotic momentum \cite{Avinash}. The semi-classical AdS/CFT correspondence therefore provides a bulk geometric realization of the stress-tensor that appears in generalized simplicity.

Once a bulk EFT is available, we can write the local gravitational EFT in first-order form as a $(d+1)$-form $L_{\rm eff}[e,\omega,\Phi]$, with $\Phi$ denoting any additional light fields, and collectively denote its fields by $\Psi=(e,\omega,\Phi)$.  Its variation takes the covariant phase-space form
\beq
\delta L_{\rm eff}={\cal E}_{\Psi}\cdot\delta\Psi+d\Theta_{\rm eff}\,.
\label{eq:EFT-first-order-variation}
\eeq
For every asymptotic conformal generator $X$, the renormalized covariant phase-space construction gives, assuming integrability, a gravitational surface-charge representative $q_X^{\rm EFT}\in\Omega^{d-1}$. After the required boundary terms and holographic counterterms are included, the finite asymptotic representative must reproduce the same charge as the boundary stress tensor,
\beq
\left[q_X^{\rm EFT}\big|_{\partial M}\right]_{\rm imp}
=
\left[{\cal J}_X[\langle T \rangle]\right]_{\rm imp}
\,.
\label{eq:EFT-stress-current}
\eeq
Equivalently, their integrals agree on every closed boundary cut.  This is the gravitational EFT realization of the usual holographic stress-tensor charge \cite{IyerWald,BalasubramanianKraus,deHaroSkenderisSolodukhin}.

When the CFT is in the gravitational EFT regime, we can combine \eqref{eq:EFT-stress-current} with the exact BF relation \eqref{eq:generalized-simplicity-cohomology}:
\beq
\left[
-\frac{1}{\kappa_{\rm BF}}\langle\zeta_X,B\rangle\Big|_{\partial M}
\right]_{\rm imp}
=
\left[q_X^{\rm EFT}\big|_{\partial M}\right]_{\rm imp}
\,.
\label{eq:generalized-simplicity-gravity-bridge}
\eeq
But note that now, the $B$ on the left is best viewed as suggesting an {\em effective} $B_{\rm grav}$ associated with the semi-classical state. By identifying a bulk EFT quantity that carries this surface current, we can make effective definitions of ``simplicity". 

For Einstein-Lovelock-type theories written in first order form, there is a natural proposal for this $B_{\rm grav}$.  When the connection enters the theory only through algebraic powers of its curvature, we can define the curvature momentum
\beq
{\cal P}_{ab}
\defi
\frac{\partial L_{\rm eff}}{\partial R^{ab}}\,.
\label{eq:curvature-momentum}
\eeq
Using $\delta R^{ab}=D(\delta\omega^{ab})$, the curvature-dependent part of the variation contributes\footnote{There are ambiguities in this, related to boundary/exact terms and to the choice of field variables \cite{IyerWald, JacobsonKangMyers}. But the statement ``GR is BF theory subject to the simplicity constraint" is usually made without reference to such refinements. So in the rest of the discussion, we will be similarly heuristic as well. We view simplicity as a fairly ad-hoc demand to obtain Einstein gravity from BF theory, and our observation is that it can be holographically motivated (not derived) as the definition of an {\em effective} $B$-field.}
\beq
\Theta_{\rm eff}^{(\omega)}
=
\delta\omega^{ab}\wedge{\cal P}_{ab}.
\label{eq:EFT-omega-symplectic-potential}
\eeq
Thus ${\cal P}_{ab}$ is the local covariant momentum conjugate to the spin connection in this first-order formulation (and it controls the corresponding Noether surface charge \cite{IyerWald}). 

But we can also consider the variation of BF theory directly. From \eqref{eq:BF-variation}, we find that
\beq
\Theta_{\rm BF}^{(\omega)}
=
\delta\omega^{ab}\wedge\frac{B_{ab}}{2\kappa_{\rm BF}}\,.
\eeq
Comparing the two suggests a natural effective-BF representative
\beq
B^{ab}_{\rm grav}
=
2\kappa_{\rm BF}{\cal P}^{ab}, \ \  \ B^{a\bullet}_{\rm grav}=0.
\label{eq:B-as-curvature-momentum}
\eeq
The factor of $2$ is fixed by our convention $\langle X,Y\rangle=\frac12X_{IJ}Y^{IJ}$.  For more general first-order EFTs containing, for example, independent torsion or derivatives of curvature, the charge-carrying quantity need not reduce to $\partial L/\partial R^{ab}$ alone. The universal statement is the charge-matching in \eqref{eq:generalized-simplicity-gravity-bridge}. The local constitutive relation \eqref{eq:B-as-curvature-momentum} applies when such a compatible BF-like polarization exists. If one instead works in metric variables or includes additional fields with non-trivial couplings to the connection, the whole discussion is somewhat moot: the simple local identification in \eqref{eq:B-as-curvature-momentum} need not take this form. We will view these as indications of the fine-tuned nature of the simplicity demand (or equivalently, the Einstein point in the EFT landscape).

For the leading two-derivative Einstein-Cartan action of Appendix~\ref{app:BF}, writing its gravitational coupling as $\kappa_{\rm grav}$,
\beq
{\cal P}^{ab}
=
\frac{1}{2\kappa_{\rm grav}(d-1)!}\,
\varepsilon^{ab}{}_{c_1\cdots c_{d-1}}\,
e^{c_1}\wedge\cdots\wedge e^{c_{d-1}}\,.
\eeq
Hence, with the common normalization $\kappa_{\rm BF}=\kappa_{\rm grav}$,
\beq
B^{ab}_{\rm grav}
=
\frac{1}{(d-1)!}\,
\varepsilon^{ab}{}_{c_1\cdots c_{d-1}}
e^{c_1}\wedge\cdots\wedge e^{c_{d-1}}\,,
\qquad
B^{a\bullet}_{\rm grav}=0\,.
\label{eq:Einstein-emergent-simplicity}
\eeq
This is precisely the gauge-fixed simplicity relation \eqref{eq:simplicity-gaugefixed}.  Ordinary simplicity is therefore the Einstein-EFT realization of the more general current-matching above. In particular, we do not view it as a condition on the universal BF sewing calculus.  Lovelock terms \cite{Lovelock} replace the Einstein expression for ${\cal P}_{ab}$ by the corresponding curvature polynomial and thereby give the natural higher-curvature generalization. 

\subsubsection{Plebanski and Spin-Foams}
\label{subsec:spin-foam-comparison}

In conventional constrained-BF gravity, simplicity is imposed as part of the definition of the gravitational sector.  In four bulk dimensions one restricts the BF two-form schematically to
\beq
B^{IJ}\ \sim\ \star_{\rm int}(e^I\wedge e^J),
\label{eq:Plebanski-simplicity-schematic}
\eeq
with higher-dimensional analogues similarly constraining $B$ to be built from a vielbein \cite{Plebanski,FreidelKrasnovPuzio}.  Spin-foam models then quantize this constrained-BF structure by implementing simplicity on representation and intertwiner data in different ways \cite{BarrettCraneRelativistic,BarrettCraneLorentzian,EnglePereiraRovelli,FreidelKrasnovNew,PerezSpinFoam}. 

Our logical order is (roughly) the reverse.  Conformal BF theory remains an unconstrained (but in a sense, ambient) sewing skeleton, while the physical CFT identifies its $B$-currents with stress-tensor Ward currents.  Only when those same data admit a suitable gravitational EFT description does a suggestive gravitational constitutive relation emerge.  Plebanski simplicity is a property of the Einstein regime in this picture, not a kinematic constraint on the bootstrap state space.

\subsubsection{Finite-Gap Corrections: Effective Simplicity}
\label{subsec:corrected-simplicity}

Once a local AdS EFT is available, departures from the Einstein limit are organized by its higher derivative and quantum expansions.  At leading order in the bulk quantum expansion, a large but finite higher-spin/single-trace gap $\Delta_{\rm gap}$ corresponds schematically to a heavy scale $M_*\ell\sim\Delta_{\rm gap}$, and integrating out heavy single-particle states gives
\beq
L_{\rm eff}
=
L_{\rm EH}
+\sum_i\frac{c_i}{M_*^{p_i}}\,
{\cal O}_i(e,R,D R,\Phi,D\Phi,\ldots)
+L_{\rm light\ matter}\,.
\label{eq:bulk-gravity-derivative-expansion}
\eeq
Generic operators that affect the gravitational surface current therefore produce corresponding departures from the Einstein simplicity relation, suppressed by powers of the heavy scale. On AdS backgrounds this is schematically an expansion in inverse powers of $M_*\ell\sim\Delta_{\rm gap}$. Under the usual large-gap and Regge assumptions, non-Einstein graviton couplings are suppressed by the higher-spin gap \cite{CamanhoEdelsteinMaldacenaZhiboedov,AfkhamiJeddiHartmanKunduTajdini,MeltzerPerlmutter}.  

The relation \eqref{eq:generalized-simplicity-gravity-bridge} can be viewed as a definition of ``effective'' $B$, in the EFT regime.  But a natural relation for a $B_{\rm grav}$ may exist only when the EFT phase space admits a compatible BF-like polarization.  For the algebraic $L(e,R)$ class above, this is precisely what happens: higher-curvature terms deform ${\cal P}_{ab}$ and hence the constitutive relation. For more general higher-derivative EFTs, however, the spin-connection symplectic potential need not have the simple $\delta\omega^{ab}\wedge{\cal P}_{ab}$ form: additional derivative variations and canonical momenta can appear. What remains invariant is the matching of the asymptotic stress-tensor current, not a universal formula for $B$. Finite $C_T$ is a separate expansion parameter: these are bulk quantum corrections and they need not be captured by a local classical constitutive relation.

\section{Comments}
\label{sec:conc}

Since we have only scratched the surface of these ideas, in lieu of a conclusion, we will make some suggestive comments. 

\paragraph{Beyond AdS.} A key observation is that the bulk sewing picture is, in principle, more general than AdS/CFT. The natural connection with spin-foam models is an indication of this. But more concretely, gluing consistency can be viewed as intrinsically bulk-based. Even though the boundary provides a privileged place in AdS/CFT, the key demand is that Wilson network amplitudes are invariant under the recoupling changes that are allowed within the sewing calculus. The latter is well-defined as a bulk property, and this offers avenues for defining theories in flat space as well as de Sitter space. 

It will be interesting to develop a sewing construction for asymptotically flat space, obtained schematically by contracting the AdS BF skeleton from the conformal group to the Poincar\'e group. BF theories are well-studied in these settings\footnote{Despite the complication that the Killing form is degenerate.} and so is harmonic analysis. In such a formulation, the BF state spaces would be organized by Poincar\'e representations and their recoupling kernels. One imagines that gluing consistency (together perhaps with some notion of unitarity) will be a natural bootstrap criterion  here. It will be very interesting if some of the pre-existing work on spin-foams and BF theory, can be refurbished for this purpose.

The case of de Sitter is also interesting, where the Wilson lines to be bootstrapped are naturally viewed as ``observers'' following \cite{WittenCrossed,CLPW,Maldacena}. In an upcoming paper \cite{CK}, we propose a bulk-bootstrap framework for dS$_3$ quantum gravity in which the Virasoro/Teichm\"uller sewing calculus \cite{PonsotTeschner1,PonsotTeschner2,TeschnerTeichmuller,CEZ1,CEZ2}, with the two chiral central charges analytically continued so that $\bar c=c^*$, is treated as the ambient skeleton, but {\em without} looking for an underlying boundary dual or complex Liouville theory \cite{CotlerJensen,CLSWorldsheet,MicroscopicdS3}. A key observation is that the sewing identities admit analytic continuation wherever the required transforms and their compositions exist: the real-$P$ spectral contour can be retained, while contour deformations and residues enter when poles are crossed \cite{CollierGobeilMaxfieldPerlmutter,RoussillonIrrational}. Schwarz reflection gives local compatibility of the continued Moore-Seiberg $F$-, $R$- and $S$-moves with an antilinear CRT involution exchanging the two conjugate chiral sectors. This suggests defining the microscopic theory in terms of a discrete spectrum of observer/Wilson-line sectors on the CRT-paired locus $\bar h=h^*$, with the spectrum and junction data constrained by gluing consistency of amplitudes defined via closed Wilson networks in $S^3$. A positivity structure on ket-bra cut kernels, preserved by conjugate-doubled sewing, provides a candidate notion of dS ``unitarity'' and makes the paired locus natural for elementary observer sectors. A nontrivial check comes from an unknotted observer loop in $S^3$: in a vacuum-cap approximation, Heegaard sewing expresses its amplitude through the vacuum row of the continued modular $S$-kernel \cite{CEZ1,CEZ2}, while CRT pairs the complete chiral contraction with its conjugate into a manifestly positive amplitude. For the vacuum observer, with a natural residue closing prescription, the logarithm of this reproduces the Gibbons-Hawking entropy together with the $-3\log S_{\rm dS}$ one-loop correction \cite{GibbonsHawking,AnninosDenefLawSun,MicroscopicdS3}. This can be viewed as a microscopically motivated refinement of the $S^3$-amplitude in pure 3D gravity, which suffers from no phase ambiguity \cite{PolchinskiPhase,Maldacena,MicroscopicdS3}. We expect much of this discussion to have natural generalizations to $d>2$ dimensions, including analytic continuation properties of BF calculus, the CRT involution, and positivity of ket-bra kernels on admissible $d$-dimensional cuts in $S^{d+1}$. The interesting question, of course, is whether a bootstrap problem framed this way has a physically interesting, non-vacuous solution space.

\paragraph{BF Fourier Transform as Higher Dimensional ``Modular" Transform.} In AdS$_3$/CFT$_2$, the Virasoro $S$-kernel has a bulk interpretation as a Fourier transform between canonically conjugate Chern-Simons variables \cite{Pradipta}.  Modular invariance itself does not generalize in this form to higher-dimensional CFTs, but the underlying polarization transform does. Indeed, for
\begin{equation}
    I_{\rm BF}
    =
    \frac{1}{2\kappa_{\rm BF}}
    \int_{M_{d+1}} B_{ab}\wedge F^{ab},
\end{equation}
the boundary symplectic potential identifies $A^{ab}$ and $B_{ab}/(2\kappa_{\rm BF})$ as conjugate variables.  The corresponding change of polarization is therefore schematically
\begin{equation}
    \widetilde\Psi[B]
    =
    \int {\cal D}A\,
    K_{\rm BF}[B,A]\,\Psi[A],
    \qquad
    K_{\rm BF}[B,A]
    \sim
    \exp\!\left[
        \frac{i}{2\kappa_{\rm BF}}
        \int_{\partial M} A^{ab}\wedge B_{ab}
    \right].
    \label{eq:BF-Fourier-outlook}
\end{equation}
We can expect this Fourier transform to have close connections to black hole physics. Let us mention a few reasons why.

As we mentioned: In the thermal-AdS filling $S^1\times B^d$, the thermal circle remains non-contractible and we can have ordinary global flat-connection holonomy. In the complementary filling $D^2\times S^{d-1}$, the same circle bounds a disk, so an undecorated flat connection would have trivial holonomy around it. Within the BF skeleton, the prescribed non-trivial boundary holonomy in this dual filling is therefore represented by a codimension-two core defect carrying the corresponding meridional monodromy. We view this as motivation for why the sewing calculus should be defect-closed. These observations have close similarities to those in \cite{Pradipta, Rajdeep}, where defects in the dual channel (Wilson lines along the core of the contractible thermal circle) were discussed.

$S^1\times S^{d-1}$ is the common boundary of $D^2\times S^{d-1}$ and $S^1\times B^d$, whose union along that boundary is topologically $S^{d+1}$. In a semiclassical gravitational regime, smooth saddles with these two topologies underlie the Hawking-Page transition \cite{HawkingPage}. For $d>2$, we need to reconcile the defect story mentioned above, with this smooth picture. This ``decoupling" between change-of-polarization and smooth-bulk-fillings is a fundamental distinction in $d>2$.

It is suggestive how this structure dovetails with  the gravity limit.  To see this, consider such a limit that admits a regular Euclidean black-hole saddle.  On a small $S^1_\perp\times H$ surrounding the horizon cross-section $H$, project the BF pairing onto its zero modes in the normal two-plane to $H$ \footnote{This is the heuristic step. The saddle is a solution of the gravity theory, not of BF theory.}. With $n^{ab}$ the corresponding binormal, define
\begin{equation}
    \Theta_H
    \equiv
    \frac12\oint_{S^1_\perp} n_{ab}A^{ab},
    \qquad
    \Pi_H
    \equiv
    \frac{1}{2\kappa_{\rm BF}}
    \int_H n_{ab}B^{ab}.
\end{equation}
The boundary pairing in \eqref{eq:BF-Fourier-outlook} then reduces in this sector to
\begin{equation}
    \frac{1}{2\kappa_{\rm BF}}
    \int_{S^1_\perp\times H} A^{ab}\wedge B_{ab}
    \;\longrightarrow\;
    \Theta_H\Pi_H,
    \qquad
    K_{\rm hor}(\Theta_H,\Pi_H)
    \sim e^{\,i\Theta_H\Pi_H}.
    \label{eq:horizon-BF-pair}
\end{equation}
Under the standard continuation of the Lorentzian normal-plane boost to a Euclidean opening angle, $\Theta_H^{\rm L}=-i\Theta_H^{\rm E}$, the Fourier factor becomes $e^{i\Theta_H^{\rm L}\Pi_H}\to e^{+\Theta_H^{\rm E}\Pi_H}$.  On the regular Euclidean gravitational saddle $\Theta_H^{\rm E}=2\pi$ and $\Pi_H$ is real, giving $\exp(2\pi\Pi_H)$.  Einstein simplicity on $B$ then leads to
\begin{equation}
    {\rm Area}(H)
    =
    \frac12\int_H n_{ab}B^{ab},
    \qquad
    \kappa_{\rm BF}=8\pi G_N,
\end{equation}
which gives
\begin{equation}
    2\pi\Pi_H
    =
    \frac{{\rm Area}(H)}{4G_N},
    \qquad
    K_{\rm hor}\big|_{\rm reg}
    \sim
    \exp\!\left[\frac{{\rm Area}(H)}{4G_N}\right].
    \label{eq:BF-area-weight}
\end{equation}
More generally, using the gravitational realization of generalized simplicity for suitable theories,
\begin{equation}
    B^{ab}_{\rm grav}
    =
    2\kappa_{\rm BF}{\cal P}^{ab},
    \qquad
    {\cal P}^{ab}
    =
    \frac{\delta L_{\rm eff}}{\delta R_{ab}},
\end{equation}
one instead finds Wald entropy
\begin{equation}
    2\pi\Pi_H
    =
    2\pi\int_H n_{ab}{\cal P}^{ab}
    =
    S_{\rm Wald},
    \qquad
    K_{\rm hor}\big|_{\rm reg}
    \sim e^{S_{\rm Wald}},
    \label{eq:BF-Wald-weight}
\end{equation}
%up to the conventional orientation sign in the definition of the binormal.
This is closely related to the classic observation of Carlip-Teitelboim that horizon area and Euclidean opening angle are canonically conjugate, and to its extension in which the conjugate momentum is the Wald entropy \cite{CarlipTeitelboim,BrusteinHadad,IyerWald}.  We emphasize that this does {\em not} amount to deriving black-hole smoothness or its microscopic density of states from unconstrained BF theory: the regular Euclidean saddle and the constitutive relation for $B$ are properties of the gravitational regime. What it shows is that the BF Fourier transform may have a role to play in this story.  It is tempting to regard this as a hint that the Cardy interpretation of the $S$-kernel row in AdS$_3$/CFT$_2$, has a broader realization  in the BF description -- presumably in terms of the high-lying density of states in the bootstrapped data. We leave a systematic development of this connection for future work \cite{Panikkar}.

\paragraph{Supersymmetry.} Supersymmetry has not played a direct role in any of our discussions. This is natural: we know in 2D that bootstrap solutions must include Liouville \cite{ZZLiouville, TeschnerLiouville}, which is an explicitly known (non-compact) CFT with only Virasoro symmetry, in the holographic $c > 1$ regime. From this perspective, supersymmetric CFTs (including ${\mathcal N}=4$ SYM and the like) are simply bootstrap solutions that exhibit extra symmetries. It is natural to suspect that if one organizes the calculation in terms of a symmetry-enhanced sewing skeleton (e.g., supersymmetric versions of BF theory in this case), we will get the corresponding enlarged multiplets directly via bootstrap. This idea was already suggested in the $d=2$ context of Chern-Simons/Virasoro in \cite{Pradipta}: it was speculated there that $PSU(1,1|2) \times PSU(1,1|2)$ Chern-Simons theory with only the spin-2 multiplet \cite{Justin} should be used as the skeleton for bootstrapping D1D5 quantum gravity. In the time since then, we \cite{Pradiptanew} have been able to explicitly show that BPS and non-BPS characters of the small ${\mathcal N}=4$ algebra can be reproduced from this skeleton by reducing the super-Chern-Simons action in a super-DS gauge, to the super-Alekseev-Shatashvili orbit action. The path integral of the latter at one loop reproduces the characters. 

More broadly, we view the fact that supersymmetry is {\em not} a part of the construction as an encouraging phenomenological hint about real world quantum gravity. Of course, it is possible that bootstrap itself may force SUSY on us in some cases (or simply rule out the existence of theories in, say, certain dimensions): but we nonetheless view this as an opening for the real world. It will also be interesting to see if there is some refinement of our proposal here that allows a distinction between holographic and non-holographic CFTs in $d>2$. Note that the $d=3$ Ising CFT \cite{Rychkov}, which is not supersymmetric, was located using crossing bootstrap. 

\paragraph{Virasoro and $d=2$.} A (not so) special case is AdS$_3$/CFT$_2$. Discussion (and various pieces of evidence) of a similar TQFT skeleton structure in $d=2$ has already appeared in \cite{Pradipta, Vishal, Rajdeep}. The key difference is that because the stress tensor is a Virasoro descendant of the vacuum (or equivalently, because of the enhancement from global conformal to Virasoro symmetry), it is more natural to organize the bootstrap in terms of the Virasoro/Teichm\"uller skeleton \cite{PonsotTeschner1}. We expect that the $d=2$ BF skeleton, with the condition that a vacuum-descendant stress tensor (and its Ward identities) is part of the spectrum, will re-organize the calculation in terms of Virasoro symmetry and will effectively lead to the Virasoro skeleton. This and some other related ideas will appear in \cite{Vishal2}. A related possibility is to enhance the symmetries to include higher spin currents, on both the bulk and the boundary.

\paragraph{Sewing Calculus vs TQFT.} Our construction suggests that the sewing calculus needed to do the bootstrap may be more fundamental than a TQFT defined on all bordisms. That the sewing conditions for 2D CFTs can be consistently formulated has been known for a long time through the remarkable work of Ponsot-Teschner \cite{PonsotTeschner1, PonsotTeschner2}, but the TQFT related to it \cite{KashaevQT, Andersen} is not a conventional Atiyah-Segal/Reshetikhin-Turaev TQFT \cite{AtiyahTQFT,ReshetikhinTuraev}. See \cite{CEZ1, CEZ2} for recent work that explores Virasoro TQFT. It may be interesting to formulate the sewing calculus itself as a fundamental mathematical object more generally.

\paragraph{Diffeomorphism Invariance as Effective Sewing Consistency.} One notable thing about the bootstrap philosophy is that the solution space need not be tied to the continuum, and in fact we expect interesting holographic solutions with discrete spectra. Diff-invariant metric theories are simply the EFT manifestations of bootstrap data when there is a stress tensor in the spectrum (along with properties like large-$N$ and large gap). This also gives a plausible interpretation for the simplicity relation in LQG/spin-foam models: rather than being imposed as a microscopic constraint on the BF skeleton, Einstein simplicity appears here as an effective constitutive relation in the Einstein regime.

\paragraph{A Mechanics for the Bulk.} Our approach in this paper can be characterized as a first attempt at a bulk re-statement of a fairly maximal version of CFT bootstrap, and then using that as the jumping off point for intrinsically bulk formulations. Clearly, this has to be made more explicit -- but the point is that there seems to be a clear path for doing this. The connections that emerged (as noted in Sections \ref{sec:asymptoticads}, \ref{sec:tower}, \ref{sec:ward-simplicity}) from this change of perspective, seem to be an indication of its naturalness. A bulk-centric approach makes the ``mechanics" of quantum gravity more transparent: bulk gravity is the effective low energy dynamics of interacting towers of Wilson lines (especially the stress tensor Wilson line) and defects. The proposal suggests that non-trivial local curvature is an effective phenomenon. Such a perspective, we hope, will be useful in understanding black holes, the Big Bang, and the nature of spacetime itself.

\section{Acknowledgments} 

The key idea in this paper -- that of viewing sewing consistency as the fundamental bulk principle underlying quantum gravity -- was outlined in \cite{Pradipta} and explored in \cite{Vishal, Rajdeep} for AdS$_3$/CFT$_2$. The new ingredient here is that we consider it in a setting where the stress tensor is an independent primary, and not a (Virasoro) descendant of the vacuum. Section \ref{sec:tower} is an adaptation of an AdS$_3$ calculation, done with Vishal Gayari. This paper has had a  long ``gestation'', and has turned into many related collaborations: I thank Apoorva Asthana, Vaibhav Burman, ChatGPT/Claude, Vishal Gayari, Raghotham Kulkarni, Rajdeep Mitra, Aditya Panikkar, Pradipta Pathak and Aravinth Ram for discussions. I have used AI as a critic/referee for vetting the proposal in Section \ref{sec:BF-bootstrap}. I take full responsibility for the final proposal and the paper.

\appendix

\section{BF Conventions and Simplicity for Euclidean AdS$_{d+1}$}\label{app:BF}

We collect the BF conventions and the gravitational specialization used in parts of the main text. The gauge group is $SO(d+1,1)$, the isometry group of Euclidean AdS$_{d+1}$ and the conformal group of its $d$-dimensional boundary. Throughout, $I,J=1,\ldots,d+2$ are vector indices with metric of signature $(+,\ldots,+,-)$. We single out the timelike index $\bullet$, and write $a,b=1,\ldots,d+1$ for the remaining Euclidean directions. Unconstrained BF theory is the topological  skeleton used in the paper. The simplicity restriction reviewed below is relevant for the nondegenerate Einstein specialization of Section~\ref{sec:ward-simplicity}.

\subsection{$SO(d+1,1)$ BF Theory}

The fields can be taken as a connection $A\in\Omega^1(M_{d+1},\mathfrak{so}(d+1,1))$ and an adjoint-valued $(d-1)$-form $B$.  Writing $A=\frac12A^{IJ}J_{IJ}$ and $B=\frac12B^{IJ}J_{IJ}$,
\beq
F^{IJ}=dA^{IJ}+A^I{}_K\wedge A^{KJ},
\qquad
(D_A X)^{IJ}=dX^{IJ}+A^I{}_K\wedge X^{KJ}+A^J{}_K\wedge X^{IK},
\label{eq:BF-basic-conventions}
\eeq
with $A^I{}_K=\eta_{KL}A^{IL}$.  We use the invariant pairing
$\langle X,Y\rangle=\frac12X_{IJ}Y^{IJ}$, so that
\eqn{eq:BF-action}{
S_{\rm BF}[A,B]
=\frac{1}{\kappa_{\rm BF}}\int_{M_{d+1}}\!\langle B\wedge F[A]\rangle
=\frac{1}{2\kappa_{\rm BF}}\int_{M_{d+1}}\!B_{IJ}\wedge F^{IJ}[A] .
}
Its variation is
\beq
\delta S_{\rm BF}
=\frac{1}{2\kappa_{\rm BF}}\int_{M_{d+1}}
\Big(\delta B_{IJ}\wedge F^{IJ}
+\delta A^{IJ}\wedge(D_A B)_{IJ}\Big)
+\frac{1}{2\kappa_{\rm BF}}\int_{\partial M_{d+1}}\delta A^{IJ}\wedge B_{IJ}.
\label{eq:BF-variation}
\eeq
Thus the boundary symplectic potential identifies $B_{IJ}/(2\kappa_{\rm BF})$ as the momentum conjugate to $A^{IJ}$ in this polarization.  Fixing $A$ at the boundary removes the displayed surface variation. Other polarizations differ by the usual boundary Legendre term.

Besides ordinary $SO(d+1,1)$ gauge transformations,
\beq
A\mapsto g^{-1}Ag+g^{-1}dg,
\qquad B\mapsto g^{-1}Bg,
\label{eq:BF-gauge}
\eeq
BF theory has the shift symmetry
\eqn{eq:BF-shift}{
\delta_\Xi A=0,
\qquad
\delta_\Xi B=D_A\Xi,
\qquad
\Xi\in\Omega^{d-2}(M_{d+1},\mathfrak{so}(d+1,1)),
}
which changes the action only by a boundary term by the Bianchi identity.  The bulk equations are
\eqn{eq:BF-eom}{
F[A]=0,
\qquad
D_A B=0.
}
Consequently there are no local bulk propagating degrees of freedom: on a contractible patch a flat $A$ is pure gauge, and in a gauge with $A=0$ the second equation gives $dB=0$, while the shift symmetry identifies $B\sim B+d\Xi$.  Global holonomies of $A$ can nevertheless survive on non-contractible cycles.

The same equations give the conserved $B$-fluxes used in the main text.  If $\Sigma_{d-1}$ is a closed oriented $(d-1)$-cycle and $\zeta$ is a globally defined adjoint section obeying $D_A\zeta=0$, then
\beq
Q_{\Sigma}[\zeta]\defi\int_{\Sigma_{d-1}}\langle\zeta,B\rangle
\label{eq:BF-flux-charge}
\eeq
is invariant under deformations of $\Sigma_{d-1}$ that do not cross sources. A generic BF flux is a charge. After the Einstein simplicity restriction it gets connected to the area (of e.g., the horizon).

\subsection{Einstein Simplicity}

For comparison with the gravitational EFT discussion, choose a unit timelike internal vector and split the connection into an $SO(d+1)$ spin connection and a coframe.  In the gauge adapted to the distinguished timelike direction we use
\eqn{eq:split-connection}{
A^{ab}\defi\omega^{ab},
\qquad
A^{a\bullet}\defi\frac{1}{\ell}\,e^a\, .
}
This uses the same AdS normalization used in Section~\ref{BRS}.  Here $\ell$ is the physical AdS radius.  The boundary conformal-Cartan decomposition of Section~\ref{sec:asymptoticads} is a different, $|1|$-graded decomposition of the same algebra. At the purely Cartan level its scale is conventional, and in the AdS specialization we choose that convention to use this same $\ell$.  With the internal signature fixed above, \eqref{eq:split-connection} gives
\beq
F^{ab}=R^{ab}[\omega]+\frac{1}{\ell^2}e^a\wedge e^b,
\qquad
F^{a\bullet}=\frac{1}{\ell}T^a,
\qquad
T^a\defi D_\omega e^a .
\label{eq:AdS-curvature-split}
\eeq
Thus $F[A]=0$ on the nondegenerate split is precisely the torsion-free locally Euclidean-AdS condition with radius $\ell$.  For $d+1>3$ this is stronger than the Einstein equations, since it selects the locally maximally symmetric AdS vacuum. In three bulk dimensions the corresponding connection flatness is locally equivalent to the vacuum Einstein equations.  In unconstrained BF theory, however, $F[A]=0$ remains the topological connection equation and is not by itself a dynamical higher-dimensional Einstein equation.

On the conventional nondegenerate gravitational branch, $B$ is restricted to a form built from coframes.  Covariantly, one may introduce one-forms $e^I$ of rank $d+1$ and a unit timelike vector $V^I$, with $V^IV_I=-1$ and $V_Ie^I=0$, and impose
\eqn{eq:simplicity-covariant}{
B^{IJ}
=\frac{1}{(d-1)!}\,
\varepsilon^{IJ}{}_{K_1\cdots K_d}\,
V^{K_1}e^{K_2}\wedge\cdots\wedge e^{K_d} .
}
In the gauge $V^I=\delta^I_{\bullet}$ this becomes
\eqn{eq:simplicity-gaugefixed}{
B^{ab}
=\frac{1}{(d-1)!}\,\varepsilon^{ab}{}_{c_1\cdots c_{d-1}}
 e^{c_1}\wedge\cdots\wedge e^{c_{d-1}},
\qquad
B^{a\bullet}=0,
}
where $\varepsilon_{a_1\cdots a_{d+1}}\defi\varepsilon_{\bullet a_1\cdots a_{d+1}}$ and $\varepsilon_{12\cdots(d+2)}=+1$.  This is the higher-dimensional continuum simplicity relation relevant for the comparison with Plebanski/spin-foam constructions. The rank condition selects the nondegenerate geometric branch.

Substituting \eqref{eq:split-connection} and \eqref{eq:simplicity-gaugefixed} into the \emph{pure} BF functional \eqref{eq:BF-action} gives
\eqn{eq:simplicity-substituted-BF}{
\left.S_{\rm BF}\right|_{\rm simp}[e,\omega]
=\frac{1}{2\kappa_{\rm BF}(d-1)!}\int_{M_{d+1}}
\varepsilon_{ab c_1\cdots c_{d-1}}\,
e^{c_1}\wedge\cdots\wedge e^{c_{d-1}}\wedge
\Big(
R^{ab}[\omega]
+\frac{1}{\ell^2}e^a\wedge e^b
\Big),
}
with $R^{ab}=d\omega^{ab}+\omega^a{}_c\wedge\omega^{cb}$. 

If \eqref{eq:simplicity-substituted-BF} were varied as an Einstein-Cartan action, its cosmological constant would instead be
\beq
\Lambda_{\rm simpBF}
=-\frac{d(d+1)}{2\ell^2},
\label{eq:Lambda-simpBF}
\eeq
so its maximally symmetric Einstein vacuum would have radius
$L_{\rm simpBF}=\ell\sqrt{(d-1)/(d+1)}$, rather than $\ell$.  Thus imposing the simple form of $B$ inside the pure $B\wedge F$ functional does \emph{not} by itself produce the gravitational EFT whose AdS vacuum is the flat connection \eqref{eq:AdS-curvature-split}.  In standard constrained-BF formulations the gravitational action contains additional non-derivative constraint/potential data: the required $B$-potential is convention- and dimension-dependent \cite{FreidelKrasnovPuzio}.

The two-derivative Einstein-Cartan EFT used in Section~\ref{sec:ward-simplicity}, with physical AdS radius $\ell$, is instead
\eqn{eq:EinsteinCartan}{
S_{\rm EC}^{(\ell)}[e,\omega]
=\frac{1}{2\kappa_{\rm grav}(d-1)!}\int_{M_{d+1}}
\varepsilon_{ab c_1\cdots c_{d-1}}\,
e^{c_1}\wedge\cdots\wedge e^{c_{d-1}}\wedge
\Big(R^{ab}[\omega]
+\frac{d-1}{d+1}\,\frac{1}{\ell^2}e^a\wedge e^b\Big).
}
Its cosmological constant is
\beq
\Lambda_{\rm AdS}
=-\frac{d(d-1)}{2\ell^2},
\label{eq:Lambda-AdS}
\eeq
and its equations give $T^a=0$ and $R_{\mu\nu}=-d\,g_{\mu\nu}/\ell^2$.  The BF-flat connection and the Einstein EFT therefore share the same AdS$_{d+1}$ vacuum geometry of radius $\ell$, even though the physical Einstein action is not obtained by the naive substitution of simplicity into the pure BF functional.  The difference is confined to the non-derivative cosmological/potential term, so the curvature momentum $\partial L_{\rm eff}/\partial R^{ab}$, and hence the simplicity/charge matching used in Section~\ref{sec:ward-simplicity}, is unchanged. 

Finally, let $\Sigma_{d-1}$ be a codimension-two surface and let $n^{ab}$ be its internal binormal, normalized by $n^{ab}n_{ab}=2$.  On the simplicity branch, \eqref{eq:simplicity-gaugefixed} gives directly
\beq
\mathrm{vol}_{\Sigma}=\frac12 n_{ab}B^{ab},
\qquad
\mathrm{Area}(\Sigma)=\frac12\int_{\Sigma_{d-1}}n_{ab}B^{ab}.
\label{eq:area-from-B}
\eeq
Thus the binormal component of $B$ measures codimension-two area in the Einstein sector.

\end{document}